\documentclass[a4paper]{amsart} 
 
\usepackage{graphicx}   
 
\newcommand{\N}{\mathbb{N}} 
\newcommand{\R}{\mathbb{R}} 
\newcommand{\B}{\mathcal{B}} 
\renewcommand{\S}{\mathcal{S}} 
\newcommand{\T}{\mathcal{T}}

\newcommand{\open}[2]{\left(#1,#2\right)} 
\newcommand{\closeopen}[2]{\left[#1,#2\right)} 
\newcommand{\openclose}[2]{\left(#1,#2\right]} 
\newcommand{\close}[2]{\left[#1,#2\right]} 
 
\newcommand{\Cardinality}{\operatorname{Card}} 
 
\newcommand{\ipotesi}[1]{\renewcommand{\theenumi}{#1}%
    \renewcommand{\labelenumi}{\textbf{\theenumi}}\item} 
 
\newtheorem{definition}{Definition}[section] 
\newtheorem{hypothesis}[definition]{Hypothesis} 
 
\newtheorem{theorem}[definition]{Theorem}

\theoremstyle{remark} 
\newtheorem{remark}[definition]{Remark}

\numberwithin{equation}{section} 
 
\begin{document} 
 
\title[A critical review of techniques for Term Structure analysis]
{A critical review of techniques for Term Structure analysis} 
 
\author[L.~Marangio, A.~Ramponi and M.~Bernaschi]{Livio Marangio, 
Alessandro Ramponi, Massimo Bernaschi} 
 
\address{Istituto per le Applicazioni del Calcolo ``Mauro Picone'' 
- C.N.R. - Via del Policlinico, 137 - 00161 Rome - ITALY} 
 
\begin{abstract} 
Fixed income markets share many features with the equity markets. 
However there are significant differences as well and many 
attempts have been done in the past to develop specific tools 
which describe (and possibly forecasts) the behavior of such 
markets. 
 
For instance, a correct pricing of fixed income securities with 
fixed cache flows requires the knowledge of the {\it term 
structure} of interest rates. A number of techniques have been 
proposed for estimating and interpreting the term structure, yet 
solid theoretical foundations and a comparative assessment of the 
results produced by these techniques are not available. 
 
In this paper we define the fundamental concepts with a mathematical
terminology. Besides that, we report about an extensive set of experiments
whose scope is to point out the strong and weak points of the most widely
used approaches in this field.

\end{abstract}
 
\maketitle 
 
\email{[livio,ramponi,massimo]@iac.rm.cnr.it} 
 
 
\setcounter{section}{-1} 
 
\section{Introduction \label{intro}}

\noindent Many ideas and techniques developed for the analysis of
financial markets can be applied both to equities and to fixed income
securities.  For instance the distribution and the dynamical evolution
of price variations are studied under the assumption that such
variations can be described by a random process, regardless of the
asset category (equity or fixed income) in exam.

However fixed income markets have many special features, thus 
specific concepts and tools have been introduced in the past to 
study (and possibly forecast) the behavior of such markets.  For 
instance, a correct pricing of fixed income securities with fixed 
cash-flows requires the knowledge of the {\it term structure of 
interest rates}. 
 
A number of ``numerical recipes'' have been proposed for 
estimating and interpreting the term structure of interest rates, 
yet a solid theoretical foundation of such concept and a 
comparative assessment of the results produced by the different 
techniques is not available. 
 
In this paper we define formally the concept of absence of 
arbitrage opportunities and derive rigorously the existence of 
discount factors. Moreover the most widely used approaches in 
estimating the term structure are presented and their performance 
is checked with an extensive set of experiments. 
 
The structure of the paper is the following:  section 
\ref{sect:fixed income securities} is a reminder of the 
terminology used in fixed income markets; section 
\ref{sect:formulazione matematica} defines a mathematical 
framework for the discount function; section \ref{sect:stima della 
term structure} describes the major techniques currently in use to 
estimate the term structure of interest rates; section 
\ref{sect:esperimenti numerici} presents the results of our 
numerical experiments; section \ref{sect:conclusioni} concludes 
the work with the future perspectives. 
%
 
\section{Fixed income securities \label{sect:fixed income securities}} 
 
\noindent A bond is a credit instrument issued by a public 
institution or a corporate company on raising a debt. When the 
size and the timing of the payments due to the investor are fully 
specified in advance, the bond falls within the {\it fixed-income 
securities} category. 
 
The main source of uncertainty for a bond markets investor is the 
default risk of the issuer, that is the chance that the issuer 
becomes unable to fulfill the commitment of paying all promised 
cash-flows on time.  For this reason the only (true) fixed-income 
securities are those issued by the governments of some developed 
countries or by ``well-known'' large corporate institutions to 
which credit rating agencies also give their highest rating: these 
are the borrowers least likely to default on their debt 
repayments. 
 
Likewise should not be regarded as fixed-income securities those 
bonds having cash-flows indexed to the rate of inflation 
(index-linked bonds) or those with embedded optionality giving 
either the issuer or the holder some discretion to redeem early or 
to convert to another security 
\footnote{To give some examples, we recall the US, UK and Japanese 
government callable bonds, giving the Treasury the option to redeem 
the bond at face value at any time between two dates specified at the 
time of issue; or the Italian government putable bonds (CTO) giving 
the holder the option to sell the bond back to the Treasury at face 
value on pre-specified dates; or the UK government and corporate 
convertible bonds giving the holder the option to convert the bond 
into another pre-specified bond at a pre-specified ratio on one or 
more pre-specified dates (see {\it e.g. 
\cite{AndersonBreedonDeaconDerryMurphy97}})} 
. 
 
The lack of any uncertainty makes fixed-income bonds a useful 
means for measuring market interest rates. 
 
An investor can purchase ``new'' bonds in the primary market ({\it 
i.e.}  directly from the issuer) or formerly issued bonds from 
other investors in the secondary market. Furthermore large 
institutional investors can sell bonds to the issuer as well, that 
is they can pay the price of the bond and undertake to repay the 
corresponding future cash-flows. Hence the bid and ask prices of a 
bond 
\footnote{respectively the price received when selling and paid 
when buying that bond} 
are set continuously, as long as it remains outstanding, by 
transactions occurring between market participants. 
 
In principle a bond can promise any pattern of future cash-flows. 
However there are two main categories. 
 
Zero-coupon bonds, known also as discount bonds, make a single 
payment at a date in the future known as the maturity date.  The 
size of this payment is called the {\it face value}, or par value, 
of the bond. The period of time up to the maturity date is the 
{\it maturity} of the bond.  US Treasury bills (Treasury 
obligations with maturity at issue of up to twelve months) take 
this form. 
 
Coupon bonds make interest payments (often referred to as coupon 
payments) of a given fraction of face value, called the {\it 
coupon rate}, at equally spaced dates up to and including the 
maturity date when the face value is also paid.  The frequency at 
which interest payments are made varies from market to market, but 
generally they are made either annually or semi-annually.  US 
Treasury notes and bonds (Treasury obligations with maturity at 
issue of twelve months up to ten years and above ten years, 
respectively) take this form.  Coupon payments on US Treasury 
notes and bonds are made every six months. 
 
In the following, for sake of simplicity, we will focus on US 
Treasury bond markets only.  There are a number of advantages in 
this choice.  First of all these markets are extremely large 
regardless of whether size is measured by outstanding or traded 
quantities, are very liquid and have uniform rules.  Moreover the 
number of issues is quite high (at present there are about two 
hundreds obligations outstanding).  These features make the study 
of such markets somewhat easier. 
 
\subsection{The Law of One Price} 
\noindent The key idea behind asset valuation in fixed-income 
markets is the absence of {\it arbitrage} opportunities, also 
known as the `Law of One Price'. 
 
The `Law of One Price' states that two portfolios of fixed-income 
securities (that is two positions taken in the marketplace through 
buying and selling fixed-income securities) that guarantee the 
investor the same future cash-flows and give him the same future 
liabilities, must sell for the same net price. Phrased another 
way, a portfolio of fixed-income securities giving the investor 
neither the right to receive any future cash-flow nor any future 
liability, must have zero initial net cost.  Conversely a 
portfolio requiring no initial net investment must give the 
investor some ``real'' liability, that is some liability whose 
size is larger than the net amount of money earned till then. 
 
If any violation of these constraints on the prices of all 
fixed-income securities outstanding at a given time occurs, then a 
profitable and risk-less investment opportunity, called an 
arbitrage opportunity, or simply an arbitrage, arises. 
 
Arbitrage opportunities may arise (sporadically) in financial 
markets but they cannot last long.  In fact, as soon as an 
arbitrage becomes known to sufficiently many investors, the prices 
will be affected as they move to take advantage of such an 
opportunity. As a consequence, prices will change and the 
arbitrage will disappear. This principle can be stated as follows: 
in an {\it efficient market} 
\footnote{we recall that according to Malkiel \cite{Malkiel92} ``A 
capital market is said to be efficient if it fully and correctly 
reflects all relevant information in determining security prices''} 
there are no permanent arbitrage opportunities. 
 
The prices of all fixed-income securities (and in particular those 
of US government bonds) as they are quoted at any given time, 
cannot, then, be independent one of the other and the existence of 
(moderate) arbitrage opportunities in financial markets can be 
viewed as a relative mispricing between correlated assets. Such 
mispricing can be ascribed to taxation, transaction costs and 
commissions associated with trading that make the net price of a 
bond different from its quoted gross price (and, hence, make the 
corresponding arbitrage opportunity not really profitable), to 
liquidity effects and other market frictions, or to 
non-synchronous quotations. 
 
\section{The mathematical setting \label{sect:formulazione matematica}} 
 
\noindent The cash-flows that a US government bond holder is 
entitled to receive are completely determined by the bond face 
value that, unless otherwise specified, will always be supposed 
equal to $\$100$, by its coupon rate and by its maturity.  For 
this reason an outstanding bond can be identified with a triplet 
$(c,m,p)$, where $c\ge 0$ is its (semi-annual) coupon rate in 
percentage points, $m>0$ is its maturity in years (computed using 
`$30/360$' convention) and $p>0$ is its {\it gross price} in 
dollars. For a US Treasury bill $c=0$ and $m\leq 1$; for a US 
Treasury note or bond $c>0$. 
 
We will state, now, formally the condition of absence of arbitrage 
opportunities and derive its consequences on the prices of US 
government bonds. This will lead us to the definition of the 
discount factors or zero-coupon bond prices. We start by giving 
some definitions. 
 
Let $\B:=\closeopen{0}{+\infty}\times\R_+\times\R_+$ be the set of 
all bonds, actually issued by the US Treasury or not, and still 
outstanding at the date in exam. 
 
Secondly, given a subset $\S$ of $\B$, denote by $T(\S)$ the 
vector of the maturities of all coupon payments of all bonds 
belonging to $\S$, sorted in increasing order.  This amounts to 
say that if $T(\S)=(t_1,\ldots,t_N)$ for some $N\in\N$, then 
\begin{equation*} 
    \{t_1,\ldots,t_N\}:=\cup_{(c,m,p)\in\S} \{m-i/2; \; 
i\in\N\cup\{0\} \;\text{and}\; m-i/2>0\} 
\end{equation*} 
and $t_i<t_{i+1}$ for every $i$. Moreover for every $(c,m,p)\in\S$ 
define the {\it cash-flows vector} $\boldsymbol{\varphi}(c,m) 
=(\varphi_1(c,m),\ldots,\varphi_N(c,m)) \in\R^N$ by 
\begin{equation*} 
    \varphi_i(c,m):= 
    \begin{cases} 
    100+c &, \,\text{if}\; m=t_i \\ 
    c     &, \,\text{if}\; t_i\in m-\frac{\N}{2} \\ 
    0     &, \,\text{otherwise} 
    \end{cases} 
    \qquad (i=1,\ldots,N). 
\end{equation*} 
Thus $\varphi_i(c,m)$ represents the cash-flow (in dollars) that 
an investor holding bond $(c,m,p)$ is entitled to receive in $t_i$ 
years' time. 
 
Thirdly, we say that a subset $\T$ of $\B$ is a {\it complete coupon 
term structure} if and only if it contains at least a bond maturing on 
each coupon payment day (see {\it e.g.} 
\cite{CampbellLoMacKinlay97}).  Observe that for a complete coupon 
term structure $\T$ the set of the maturities of all coupon payments 
of all bonds belonging to $\T$ is the same as the set of the 
maturities of all bonds belonging to $\T$, that is, if 
$T(\T)=(t_1,\ldots,t_N)$, then 
 
\begin{equation*} 
    \{t_1,\ldots,t_N\}:=\cup_{(c,m,p)\in\T} \{m\}. 
\end{equation*} 
 
Finally we represent a {\it portfolio} of $B$ bonds 
$(c_1,m_1,p_1),\ldots,(c_B,m_B,p_B)$ by a vector 
$(q_1,\ldots,q_B)\in\R^B$ such that for each $j$, $q_j$ represents 
the traded quantity of bond $(c_j,m_j,p_j)$. By definition, 
$q_j>0$ means that the investor has bought $\$ 100 q_j$ face value 
of bond $j$ whereas $q_j<0$ means that he has sold $\$ 100 |q_j|$ 
face value of that bond. 
 
We are, now, in a position to state formally the condition of absence 
of arbitrage opportunities among the bonds belonging to a given set 
$\S\subset\B$, and to show which constraints on their prices such 
condition implies (see Theorem \ref{thm:fattori di sconto} below). 
 
\begin{definition} 
    Let $\S$ be a subset of $\B$ and let $T(\S)=(t_1,\ldots,t_N)$ 
for some $N\in\N$. We say that $\S$ satisfies the hypothesis of 
{\it Absence of Arbitrage Opportunities} if and only if, given 
$(c_1,m_1,p_1),\ldots,(c_B,m_B,p_B)\in\S$ ($B\in\N$) the following 
three conditions are fulfilled: 
\begin{enumerate} 
    \ipotesi{$(NA_1)$}\label{cond:arbitraggio 1} 
if $(q_1,\ldots,q_B)\in\R^B$ is such that $\sum_{j=1}^B q_j 
\boldsymbol{\varphi}(c_j,m_j)=\boldsymbol{0}$, then $\sum_{j=1}^B q_j 
p_j=0$; 
    \ipotesi{$(NA_2)$}\label{cond:arbitraggio 2} 
if $(q_1,\ldots,q_B)\in\R^B$ is such that 
\begin{equation*} 
    \sum_{j=1}^B q_j \varphi_i(c_j,m_j) = 
    \begin{cases} 
        f_{\bar{\imath}} &, \quad\text{if}\; i =\bar{\imath} \\ 
        0                &, \quad\text{otherwise} 
    \end{cases}, 
\end{equation*} 
for some $f_{\bar{\imath}} >0$, then $0 <\sum_{j=1}^B q_j p_j 
<f_{\bar{\imath}}$; 
    \ipotesi{$(NA_3)$}\label{cond:arbitraggio 3} 
if $(q_1,\ldots,q_B)\in\R^B\backslash\{0\}$ is such that 
$\sum_{j=1}^B q_j p_j=0$, then there exists 
$\bar{\imath}\in\{1,\ldots,N\}$ such that 
$\sum_{i=1}^{\bar{\imath}} \sum_{j=1}^B q_j \varphi_i(c_j,m_j) 
<0$. 
\end{enumerate} 
\end{definition} 
 
\begin{remark} 
    The $i$th component of the vector 
$\sum_{j=1}^B q_j\boldsymbol{\varphi}(c_j,m_j)$ represents, if it 
is positive, the total cash-flow that an investor holding 
portfolio $(q_1,\ldots,q_B)$ will receive in $t_i$ years' time; if 
it is negative, the total liability that he will have at that 
date.  The quantity $\sum_{j=1}^B q_j p_j$ is the (net) price of 
the portfolio. 
 
Hence, condition \ref{cond:arbitraggio 1} means that a portfolio 
of bonds giving the investor neither the right to receive any 
future cash-flow nor any future liability must have zero initial 
net cost.  Conversely \ref{cond:arbitraggio 3} means that a 
non-trivial portfolio requiring no initial net investment must, 
sooner or later, give the investor a liability of larger size than 
the total net amount of money received till then. 
 
The meaning of condition \ref{cond:arbitraggio 2} is self-evident and 
will not be discussed any further. 
\end{remark} 
 
\begin{definition} 
    Given $B+1$ bonds $(c,m,p),(c_1,m_1,p_1),\ldots,(c_B,m_B,p_B)$ 
in $\B$, a portfolio $(q_1,\ldots,q_B)$ of the last $B$ bonds such 
that $\sum_{j=1}^B 
q_j\boldsymbol{\varphi}(c_j,m_j)=\boldsymbol{\varphi}(c,m)$, is 
called a {\it replicating portfolio} of bond $(c,m,p)$. In fact it 
entitles an investor to receive exactly the same cash-flows as if 
he held bond $(c,m,p)$. If, furthermore, each bond $(c_j,m_j,p_j)$ 
expires on a different coupon payment date of bond $(c,m,p)$, then 
we will refer to such a portfolio as a minimal replicating 
portfolio. 
\end{definition} 
 
\begin{remark} 
    Since the case $q_j\equiv 0$ is trivial, condition 
\ref{cond:arbitraggio 1} amounts to say that the price of each 
bond $(c,m,p)$ in $\S$ equals the ones of all its replicating 
portfolios, whenever they exist. 
\end{remark} 
 
\begin{theorem}\label{thm:fattori di sconto} 
    Let $\T\subset\B$ be a complete coupon term structure and let 
$T(\T)=(t_1,\ldots,t_N)$ for some $N\in\N$.  If $\Cardinality(\T) 
\geq N$ and if condition \ref{cond:arbitraggio 1} is satisfied, 
then there exist $d_1,\ldots,d_N\in\R$ such that for every 
$(c,m,p)\in\T$ 
\begin{equation}\label{eq:prezzo} 
    p =\sum_{i=1}^N d_i \varphi_i(c,m). 
\end{equation} 
If, furthermore, conditions \ref{cond:arbitraggio 2} and 
\ref{cond:arbitraggio 3} are fulfilled as well, then 
$1>d_1>d_2>\ldots>d_N>0$. 
\end{theorem} 
 
A proof of this theorem is reported in the Appendix. 
 
\begin{remark}\label{rem:completa} 
    In order to prove Theorem \ref{thm:fattori di sconto} we need 
to find $N$ bonds in $\T$ such that the matrix whose columns are 
their cash-flows vectors have maximum rank $N$. The hypothesis 
that $\T$ forms a complete coupon term structure just helps to 
ensure that this choice is possible. However this is not the only 
case in which this is true. To realize this it is sufficient to 
think that if $\S=\{(c_1,m_1,p_1),(c_2,m_2,p_2),(c_3,m_3,p_3)\}$ 
with $m_1=\frac{1}{2}$, $m_2=m_3=\frac{3}{2}$ and (of course) $c_2 
\neq c_3$, then the matrix whose columns are the cash-flows 
vectors of these three bonds has maximum rank $3$, even if $\S$ is 
not a complete coupon term structure. 
\end{remark} 
 
\begin{remark} 
    By equation \eqref{eq:prezzo} with $c=0$ it is apparent that 
each $d_i$ represents the price of a (hypothetical) zero-coupon 
bond in $\T$ with unitary face value and maturity $m=t_i$.  For 
this reason $d_1,\ldots,d_N$ are called the {\it discount 
factors}, or zero-coupon bond prices, corresponding to the 
maturities $t_1,\ldots,t_N$. 
\end{remark} 
 
\begin{definition} 
    Given $N$ discount factors $d_1,\ldots,d_N$ corresponding to 
maturities $t_1,\ldots,t_N$, we define the {\it spot rate} (of 
return), or zero-coupon yield, $s_i$, by 
\begin{equation*} 
    s_i:=-\frac{1}{t_i}\lg d_i 
\end{equation*} 
Furthermore, given $P\in\N$, if there exists $j(i)>i$ such that 
$t_{j(i)}-t_i=\frac{P}{2}$, we define the {\it $P$-periods forward 
rate} (of return) 
\footnote{as it is common in financial literature, we will refer 
to any six months' time interval as a period} 
, $f^P_i$, by 
\begin{equation*} 
    f^P_i:=-\frac{2}{P}\lg\frac{d_{j(i)}}{d_i}. 
\end{equation*} 
\end{definition} 
 
Since $\frac{1}{d_i}=\exp\left(\int_0^{t_i} s_i \, dt\right)$, 
then $s_i$ represents the continuously compounded rate of return 
per unit time on a $t_i$-years investment, made today, in the bond 
markets. Analogously $f^P_i$ represents the continuously 
compounded rate of return per unit time on a $P$-periods 
investment to be made $t_i$ years ahead. 
 
As it is common in financial literature, we will refer to the set 
of the spot rates corresponding to all maturities as the 
(cross-sectional) {\it term structure of interest rates}. 
 
\subsection{The method of Carleton and Cooper} 
\noindent If a complete coupon term structure is available and if 
conditions \ref{cond:arbitraggio 1}-\ref{cond:arbitraggio 3} are 
(strictly) satisfied, then Theorem \ref{thm:fattori di sconto} 
tells us how to compute a complete set of decreasing discount 
factors (one for each coupon payment date) that summarize the 
whole information carried about nominal discount rates by bonds in 
our data set. 
 
When small violations of condition \ref{cond:arbitraggio 1} are 
detected in the bond set $\T$ forming a complete coupon term 
structure, Carleton and Cooper \cite{CarletonCooper76} suggested 
to consider the prices of the bonds in $\T$ as random variables 
with expected values satisfying \ref{cond:arbitraggio 1} - 
\ref{cond:arbitraggio 3}. According to their hint, equation 
\eqref{eq:prezzo} must be modified by adding a bond-specific error 
term to (say) its right-hand side: 
\begin{equation}\label{eq:Carleton} 
    p_j=\sum_{i=1}^N d_i\varphi_i(c_j,m_j)+\varepsilon_j, 
\end{equation} 
where $j$ ranges over all bonds in the set $\T$, the 
$\varepsilon_j$'s are random variables with 
$E\{\varepsilon_j\}\equiv 0$, and $1>d_1>d_2>\ldots>d_N>0$. 
 
Estimators, $\hat{d}_1,\ldots,\hat{d}_N$, of the discount factors 
$d_1,\ldots,d_N$, subjected to the condition 
$1>\hat{d}_1>\hat{d}_2>\ldots>\hat{d}_N>0$, can, then, be attained 
by a constrained least squares procedure with all bonds in the set 
$\T$. 
 
We recall, however, that the least-squares estimators 
$\hat{d}_1,\ldots,\hat{d}_N$ of the $d_i$'s exist and are unique 
provided that the regressors' matrix, 
$\Phi=((\varphi_i(c_j,m_j)))$, has maximum rank $N$, which is 
certainly true if $\T$ is a complete coupon term structure (see 
Remark \ref{rem:completa}). If, on the contrary, the rank of the 
matrix $\Phi$ is less than $N$, then the problem of estimating the 
discount factors $d_1,\ldots,d_N$ by a least squares procedure is 
improperly posed, or ill-posed, in the sense of Hadamard (see, 
{\it e.g.}, \cite{Kirsch96}) and does not admit any solution. 
Hence the completeness hypothesis of the term structure of $\T$ 
(or a weaker version of its) is necessary to apply the method of 
Carleton and Cooper. 
 
\section{Estimation of the term structure \label{sect:stima della term structure}} 
 
In general, the bond set you decide to use for your analysis does 
not form a complete coupon term structure. This is true, for 
example, for the set we used in the present paper. To realize this 
look at the next figure \ref{fig:spettro}, where we have reported 
the maturity spectrum of all of $206$ bonds in our sample. 
 
\begin{figure} 
\begin{center} 
\includegraphics*[scale=0.8]{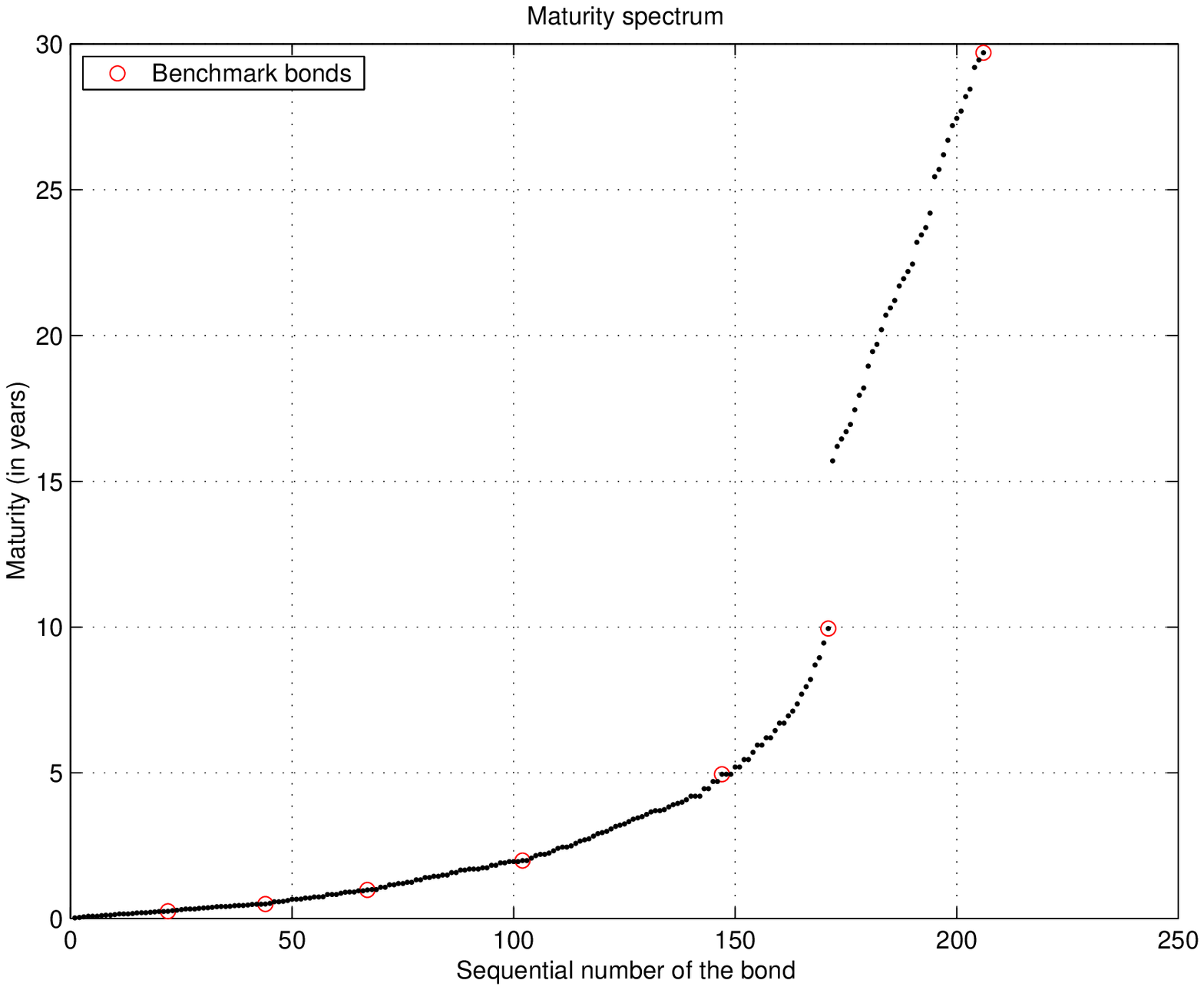} 
\end{center} 
\caption{\label{fig:spettro}} 
\end{figure} 
 
Since no bond has maturity of $10$ up to $15$ years and the 
maturity of the longest (coupon) bond reaches about $30$ years, 
the term structure of coupon bonds in our data set cannot be 
complete. 
 
In $1971$ McCulloch \cite{McCulloch71} introduced a new technique 
for studying the term structure of interest rates based on the 
following assumption 
\footnote{Actually McCulloch's statement of Hypothesis 
\ref{hyp:funzione di sconto} differs somewhat from ours: he 
assumed that coupon payments arrive in a continuous stream instead 
of semiannually. However our formulation corresponds to the one 
preferred by most authors after McCulloch} 
. 
 
\begin{hypothesis}\label{hyp:funzione di sconto} 
    If the bond set $\S\subset\B$ satisfies conditions 
\ref{cond:arbitraggio 1}-\ref{cond:arbitraggio 3}, then there 
exists a decreasing function 
$d\colon\closeopen{0}{+\infty}\to\openclose{0}{1}$ such that 
$d(0)=1$, $\lim_{t\to +\infty} d(t)=0$ and that for every bond 
$(c,m,p)\in\S$ 
\begin{equation*} 
    p=\sum_{i=1}^N d(t_i)\varphi_i(c,m). 
\end{equation*} 
\end{hypothesis} 
 
\begin{remark} 
    Function $d$ is called the {\it discount function} for the 
bond set $\S$.  For each $t\geq 0$, $d(t)$ represents, then, the 
present value of $\$ 1$ to be received in $t$ years' time.  When 
$\S=\T$ forms a complete coupon term structure, the discount 
factors $d_1,\ldots,d_N$, whose existence is guaranteed by Theorem 
\ref{thm:fattori di sconto}, are just the values that $d$ takes on 
the corresponding maturities $t_1,\ldots,t_N$, that is 
$d_i=d(t_i)$ for every $i$. 
\end{remark} 
 
The definitions of spot rate and of $P$-periods forward rate can 
be easily extended, by means of the discount function $d$, to all 
maturities $t\geq 0$. In particular the spot rate function 
$s\colon\open{0}{+\infty}\to\R_+$ is defined by 
\begin{equation}\label{eq:spot continua} 
    s(t):=-\frac{1}{t}\lg d(t) 
\end{equation} 
and the $P$-periods forward rate function 
$f^P\colon\closeopen{0}{+\infty}\to\R_+$ by 
\begin{equation*} 
    f^P(t):=-\frac{2}{P}\lg\frac{d(t+P/2)}{d(t)}. 
\end{equation*} 
 
Furthermore, if $d$ is differentiable, we can define the {\it 
instantaneous forward rate} (of return) function 
$f\colon\closeopen{0}{+\infty}\to\R_+$ by 
\begin{equation}\label{eq:forward continua} 
    f(t):=-\frac{d}{dt}\lg d(t). 
\end{equation} 
 
Then $f(t)$ represents the simple net rate of return per unit time 
of an investment in the bond markets over an infinitesimal time 
interval starting $t$ years ahead. Moreover the $P$-periods 
forward rate $f^P(t)$ equals the average of the instantaneous 
forward rate $f$ over the length $P/2$ interval 
$\close{t}{t+P/2}$, that is 
\begin{equation*} 
    f^P(t)\equiv\frac{2}{P}\int_t^{t+P/2} f(s) \, ds. 
\end{equation*} 
 
It is important to understand, however, that the existence of a 
discount function, even when conditions \ref{cond:arbitraggio 
1}-\ref{cond:arbitraggio 3} are strictly satisfied, is nothing but a 
hypothesis that scholars and practitioners assume in order to fill the 
gaps in the maturity spectrum of bond samples that do not form a 
complete coupon term structure.  It is by no means a consequence of 
\ref{cond:arbitraggio 1}-\ref{cond:arbitraggio 3}.  As we have already 
pointed out at the end of section \ref{sect:formulazione matematica}, 
building a complete set of discount factors is an ill-conditioned 
problem when the term structure of coupon bonds in the sample is 
incomplete.  This explains the difficulties met with by people when 
trying to estimate a discount function from a generic bond data set. 
 
As to the regularity properties of the discount function, some 
researchers (see, {\it e.g.}, \cite{ColemanFisherIbbotson92}) 
assert that on the ground of mere economic considerations we can 
only require that $d$ be monotonic decreasing, any further 
restrictions being not justified.  On the contrary, most authors 
assume also that $d$ be twice differentiable so attaining a smooth 
forward rate curve. In particular Langetieg and Smoot 
\cite{LangetiegSmoot89} argue that a non-smooth forward rate 
function $f$ should give rise to arbitrage opportunities and that, 
consequently, any irregularity of $f$ should be quickly priced out 
in an efficient market. 
 
If we assume Hypothesis \ref{hyp:funzione di sconto}, then the 
problem of estimating the term structure of interest rates from a 
bond data set reduces to choosing an $n$-parameter family of 
functions 
$d(\cdot;\boldsymbol{\alpha})\colon\closeopen{0}{+\infty}\to\R$ 
that we consider capable of capturing the main characteristics of 
the bond markets in exam and determining the parameter vector 
$\boldsymbol{\alpha}\in\R^n$ by minimizing the sum of squared 
residuals (least squares fitting procedure) 
\begin{equation}\label{eq:minimizzazione} 
    \min_{\boldsymbol{\alpha}} \sum_j \left|p_j -\sum_{i=1}^N 
d(t_i;\boldsymbol{\alpha})\varphi_i(c_j,m_j)\right|^2, 
\end{equation} 
where $j$ ranges over all bonds $(c_j,m_j,p_j)$ in our data set. 
The choice of the functions $d(\cdot;\boldsymbol{\alpha})$, that 
at the very end is always a matter of judgment, is crucial to the 
quality of our fit. 
 
The test functions $d(\cdot;\boldsymbol{\alpha})$ that are usually 
employed in the fitting procedure \eqref{eq:minimizzazione} may be 
roughly divided in two categories: piecewise (polynomial or 
exponential) functions or {\it spline} and functions generated by 
``parsimonious'' models. 
 
Spline functions are more capable of capturing genuine bends in 
the discount function but have the drawback of over-fitting risk; 
parsimonious models, on the contrary, are less flexible but do a 
better job in removing noise from the data. So the choice of 
spline functions is better for getting a ``good fit'' whereas 
parsimonious models should be privileged for sake of 
``smoothness''. 
 
\subsection{Spline based techniques \label{subsec:metodi spline}} 
\noindent The most widely used spline method, specially among 
practitioners, is due to McCulloch \cite{McCulloch71} (see also 
\cite{McCulloch75}) and is based on the use of polynomial spline 
functions. 
 
We recall that, given $-\infty<K_1<K_2<\ldots<K_k<+\infty$ (for 
some $k>2$) and $r\in\N\cup\{0\}$, we say that 
$f\colon\close{K_1}{K_k}\to\R$ is an $r$-degree polynomial spline 
function with knot points $K_1,K_2,\ldots,K_k$, if and only if $f$ 
is an $r$-degree polynomial in each interval 
$\close{K_i}{K_{i+1}}$ and if it is continuous, along with all its 
derivatives up to order $r-1$ (if $r>0$), in $\close{K_1}{K_k}$. 
 
The reasons underlying the choice of polynomial spline functions 
are as follows.  It is apparent that a very simple, but somehow 
na\"{\i}ve, approach to the problem of fitting a discount function 
to a bond data sample is to use polynomial test functions. However 
such functions have uniform resolving power, that is they tend to 
fit better at the short end of the maturity spectrum where the 
greatest concentration of bond maturities occurs (see figure 
\ref{fig:spettro}), and worse at the long end. A possible way out 
is, of course, to increase the order of the polynomial but this 
can cause instability in the parameter estimates. On the other 
hand, practitioners often require methods endowed with a 
sufficiently high flexibility to have the chance to choose {\it a 
priori} how good their fit of the discount function will be in 
various regions of the maturity spectrum. For instance, they may 
desire to improve the goodness of their fit in those maturity 
ranges where many bonds are traded. 
 
A fitting procedure that meets all these requirements involves the 
use of polynomial spline functions. By means of a convenient 
choice of the number and the position of the knots, such method 
allows to achieve the desired resolution degree along the whole 
maturity spectrum, even employing relatively low order polynomials 
(which means more stable curves). 
 
So the choice of the number and the position of knot points 
affects dramatically the quality of the estimated term structure 
of interest rates. 
 
McCulloch \cite{McCulloch71} suggested to choose a number $k$ of 
knots equal (approximately) to the square root of bonds in the 
sample and to place them in such a way that $K_1=0$, $K_k$ equals 
the maturity of the longest bond and that each subinterval 
$\close{K_i}{K_{i+1}}$ contains approximately the same number of 
observed maturities. This choice should provide an increasing 
resolving power as the number of observations increases and should 
avoid over-fitting phenomena. 
 
Currently, practitioners and financial analysts prefer to place 
knots at $0$, $1$, $3$, $5$, $7$, $11=10^+$ and $30$ years, so 
splitting up the whole maturity range into the intervals 
$\open{0}{1}$, $\open{1}{3}$, $\open{3}{5}$, $\open{5}{7}$, 
$\open{7}{11}$, and $\open{11}{30}$. The rationale behind this 
choice is that most asset management companies divide their assets 
into groups (or categories) that always lie in one of the above 
intervals of the maturity spectrum. This fact is reflected also in 
the existence of several benchmark securities with maturity of 
about $1$, $3$, $5$, $7$, $10$ and $30$ years. 
 
Given $k$ knot points $K_1,\ldots,K_k$, the set of all $r$-degree 
polynomial spline functions on $\close{K_1}{K_k}$ form a linear space 
of dimension $k+r-1$.  As a matter of fact we have the chance of 
choosing arbitrarily the $r+1$ coefficients of the polynomial in the 
first subinterval $\close{K_1}{K_2}$ and one coefficient for each 
polynomial in the $k-2$ remaining subintervals $\close{K_i}{K_{i+1}}$ 
with $i=2,\ldots,k-1$.  The others parameters are fixed by the 
conditions $f^{(n)}(K_i-0)=f^{(n)}(K_i+0)$ for all 
$i\in\{2,\ldots,k-1\}$ and all $n\in\{0,\ldots,r-1\}$. 
 
Let $\{f_0,f_1,\ldots,f_{r+k-2}\}$ be a base for the linear space of 
$r$-degree polynomial spline functions with knot points 
$K_1,\ldots,K_k$.  Without loss of generality we can assume that 
$f_0(t)\equiv 1$ (as a constant function is a polynomial spline 
function) and that $f_1(0)=\ldots=f_{r+k-2}(0)=0$.  Then 
$\boldsymbol{\alpha}=(\alpha_1,\ldots,\alpha_{r+k-2})$ and the 
McCulloch model for the discount function is 
\begin{equation*} 
    d(t;\boldsymbol{\alpha})=1+\sum_{j=1}^{r+k-2}\alpha_j f_j(t). 
\end{equation*} 
The corresponding least squares problem is therefore linear in the 
parameter vector $\boldsymbol{\alpha}$. 
 
A major limitation of this method is that it does not force a 
monotonous decreasing behavior of the resulting discount function.  As 
a consequence, meaningless results, like negative forward 
rates, may be obtained.  In this respect 
Barzanti and Corradi \cite{BarzantiCorradixx.2} (see also 
\cite{BarzantiCorradi98} and \cite{BarzantiCorradixx.2}) have recently, 
proposed the introduction a set of linear constraints on the parameters 
$\alpha_1,\ldots,\alpha_{r+k-2}$ that ensure the desired monotonicity 
of the employed spline functions. 
 
A further criticism directed by Vasicek and Fong 
\cite{VasicekFong82} to McCulloch's method is that most 
general-equilibrium models of the term structure of interest rates 
(see, {\it e.g.}, \cite{Vasicek77} and \cite{CoxIngersollRoss85a}) 
expect an exponential form for the discount function.  However 
piecewise polynomials have a different curvature compared to an 
exponential function. Hence they conclude that ``conventional'' 
polynomial splines cannot provide a good fit to an 
exponential-like discount function. 
 
Following the arguments of Vasicek and Fong, Langetieg and Smoot 
\cite{LangetiegSmoot89} and Coleman, Fisher and Ibbotson 
\cite{ColemanFisherIbbotson92} proposed, respectively, to fit a 
polynomial spline to the spot rate function and to the 
instantaneous forward rate function rather than to the discount 
function 
\footnote{Actually Coleman, Fisher and Ibbotson proposed only the 
use of piecewise constant functions, that is of $0$-degree 
polynomial splines. However their method can be easily extended to 
polynomial splines of any order.} 
. They argued that, assuming an exponential form for the discount 
function, their regression procedure should give better results 
({\it i.e.} more ``realistic'' from the financial viewpoint) 
compared to McCulloch's technique. 
 
By \eqref{eq:spot continua} and \eqref{eq:forward continua} 
$d(t)\equiv\exp(-t s(t))$ and $d(t)\equiv\exp\left(-\int_0^t f(s) 
\, ds\right)$. Hence the method by Langetieg and Smoot amounts to 
take $\boldsymbol{\alpha}=(\alpha_0,\ldots,\alpha_{r+k-2})$ (for 
some $r\in\N\cup\{0\}$ and some $k\geq 2$) and 
\begin{equation*} 
    d(t;\boldsymbol{\alpha}) 
=\exp\left(t \sum_{j=0}^{r+k-2}\alpha_j f_j(t)\right), 
\end{equation*} 
where $\{f_0,f_1,\ldots,f_{r+k-2}\}$ is a base for the 
linear space of $r$-degree polynomial spline functions with knot 
points $K_1,\ldots,K_k$. 
 
Analogously, since the integral of an $r$-degree polynomial spline 
function is an $(r+1)$-degree polynomial spline, the method by 
Coleman, Fisher and Ibbotson amounts to take 
$\boldsymbol{\alpha}=(\alpha_0,\ldots,\alpha_{r+k-1})$ and 
\begin{equation*} 
    d(t;\boldsymbol{\alpha}) 
=\exp\left(1 +\sum_{j=1}^{r+k-1}\alpha_j f_j(t)\right), 
\end{equation*} 
where, as above, $\{f_0,f_1,\ldots,f_{r+k-1}\}$ is a base for the 
linear space of $(r+1)$-degree polynomial spline functions with 
knot points $K_1,\ldots,K_k$ and with the property that 
$f_0(t)\equiv 1$ and that $f_1(0)=\ldots=f_{r+k-1}(0)=0$. 
 
The methods by Langetieg and Smoot and by Coleman, Fisher and 
Ibbotson share the same limitation of the method by McCulloch in 
that it is impossible to force a monotonous decreasing behavior 
of the resulting discount function by simply introducing linear 
constraints on the parameters to be determined. The only exception 
is for the method by Coleman, Fisher and Ibbotson with $r=0$, that 
is when a piecewise constant functions are used to fit the 
instantaneous forward rate function. Hence, again, financially 
unacceptable results may be found. 
 
\subsection{Parsimonious models} 
\noindent It is readily apparent that, although spline based 
techniques are still widely used to estimate the term structure of 
interest rates, they present many drawbacks (as pointed out, for 
example, by Shea \cite{Shea84,Shea85}).  As a matter of fact 
several choices, such as the number and the position of knots and 
the degree of the basic polynomials, affect dramatically the 
quality of fit. 
 
An alternative approach to the estimation of the term structure of 
interest rates is the use of the so-called {\it parsimonious} models. 
The main idea behind this class of models is to postulate a unique 
functional form on the whole range of maturities for the discount 
function.  This implies that in parsimonious models some basic 
properties of the discount function, such as monotonicity, we are 
interested in can be imposed {\it a priori}.  Furthermore, the number 
of parameters to estimate is typically much smaller than that needed 
for a typical spline based approach. 
 
Let us describe briefly some of the parsimonious models proposed 
in the literature. 
 
In \cite{ChambersCarletonWaldman84} the functional form for the 
discount function is proposed to be 
\begin{equation*} 
    d(t;\boldsymbol{\alpha}) =\exp(-\sum_{j=1}^n \alpha_j t^j). 
\end{equation*} 
 
The unknown parameters $\alpha_1,\ldots,\alpha_n$ can be estimated 
either by (non-linear) least square regression or by a maximum 
likelihood approach.  The latter one is found to perform better, 
probably, because of the explicit consideration of the 
heteroskedasticity in the data.  Notice that this model is a 
particular case of the ones by Langetieg and Smoot and by Coleman, 
Fisher and Ibbotson with just two knots. 
 
A particularly relevant approach is the one proposed by Nelson and 
Siegel \cite{NelsonSiegel87}.  They attempted to model explicitly the 
implied instantaneous forward rate function $f$.  The Expectation 
Hypothesis provides a heuristic motivation for their model since $f$ 
is modeled as the solution of a second order differential equation. 
The functional form suggested for $f$ is 
\begin{equation}\label{eq:Nelson} 
    f(t;\boldsymbol{\alpha}) = \beta_0 +\beta_1 e^{-t/\tau} +\beta_2 
\frac{t}{\tau} e^{-t/\tau}, 
\end{equation} 
where $\boldsymbol{\alpha}=(\beta_0,\beta_1,\beta_2,\tau)$. It 
implies the following model for the discount function: 
\begin{equation*} 
    d(t;\boldsymbol{\alpha}) =\exp\left(-t \left(\beta_0 +(\beta_1 
+\beta_2)\frac{\tau}{t}\left(1 -e^{-t/\tau}\right) -\beta_2 
e^{-t/\tau}\right)\right). 
\end{equation*} 
 
The flexibility of this model is explained by the authors by observing 
that the instantaneous forward rate is the linear combination of three 
components, $f_s(t):=e^{-t/\tau}$, $f_m(t):=\tau e^{-t/\tau}$ and 
$f_l(t):=1$, modeling respectively the short-, medium-, and long-term 
behavior of $f$.  The parameters $\beta_0$, $\beta_1$ and $\beta_2$ 
measure respectively the strength of each component, whereas $\tau$ is 
a time-scale factor. 
 
In \cite{Svensson94} an extended model for the instantaneous forward 
rate $f$ was proposed to increase the flexibility of the original 
model by Nelson and Siegel model.  It was obtained by adding a fourth 
term in equation \eqref{eq:Nelson}, with two extra parameters, which 
allow the instantaneous forward rate curve for a second ``hump'': 
\begin{equation*} 
   f(t;\boldsymbol{\alpha}) =\beta_0 +\beta_1 e^{-t/\tau_1} +\beta_2 
\frac{t}{\tau_1} e^{-t/\tau_1} +\beta_3 \frac{t}{\tau_2} 
e^{-t/\tau_2}. 
\end{equation*} 
 
The model proposed by a J.~P.~Morgan group and reported by Wiseman 
\cite{Wiseman94}, is the so-called Exponential Model.  The 
instantaneous forward rate function is modeled by 
\begin{equation*} 
    f(t;\boldsymbol{\alpha})=\sum_{j=0}^n a_j \exp (-b_j t), 
\end{equation*} 
where $a_1,b_1,\ldots,a_n,b_n$ are the parameters to be estimated.  It 
has been noticed that this model is able to capture the macro shape of 
the forward rate curve rather than the features of a single data set. 
 
\section{Numerical Experiments \label{sect:esperimenti numerici}} 
 
\noindent In this section we report the results of a set of numerical 
experiments for estimating the term structure of interest rates from 
a cross-sectional data set of US government bonds. 
 
The data set was made available by Datastream.  It contains the 
annual coupon rate, the maturity date and the gross (or dirty or 
tel quel) price 
\footnote{{\it i.e.} the sum of the quoted, or clean, price and 
the accrued interest (computed by the '$30/360$' convention)} 
for all (fixed-income) US Treasury bills, notes and bonds outstanding 
at the dates from May $31^{\text{st}}$ $1999$ to June $11^{\text{th}}$ 
$1999$ (two weeks). 
 
Hereafter, for economy of presentation, since the results are very 
similar, only the results for June $3^{\text{rd}}$ $1999$ (when 
$33$ bills and $173$ notes and bonds were outstanding) will be 
presented and analyzed. 
 
\begin{figure} 
\begin{center} 
\includegraphics*[scale=0.8]{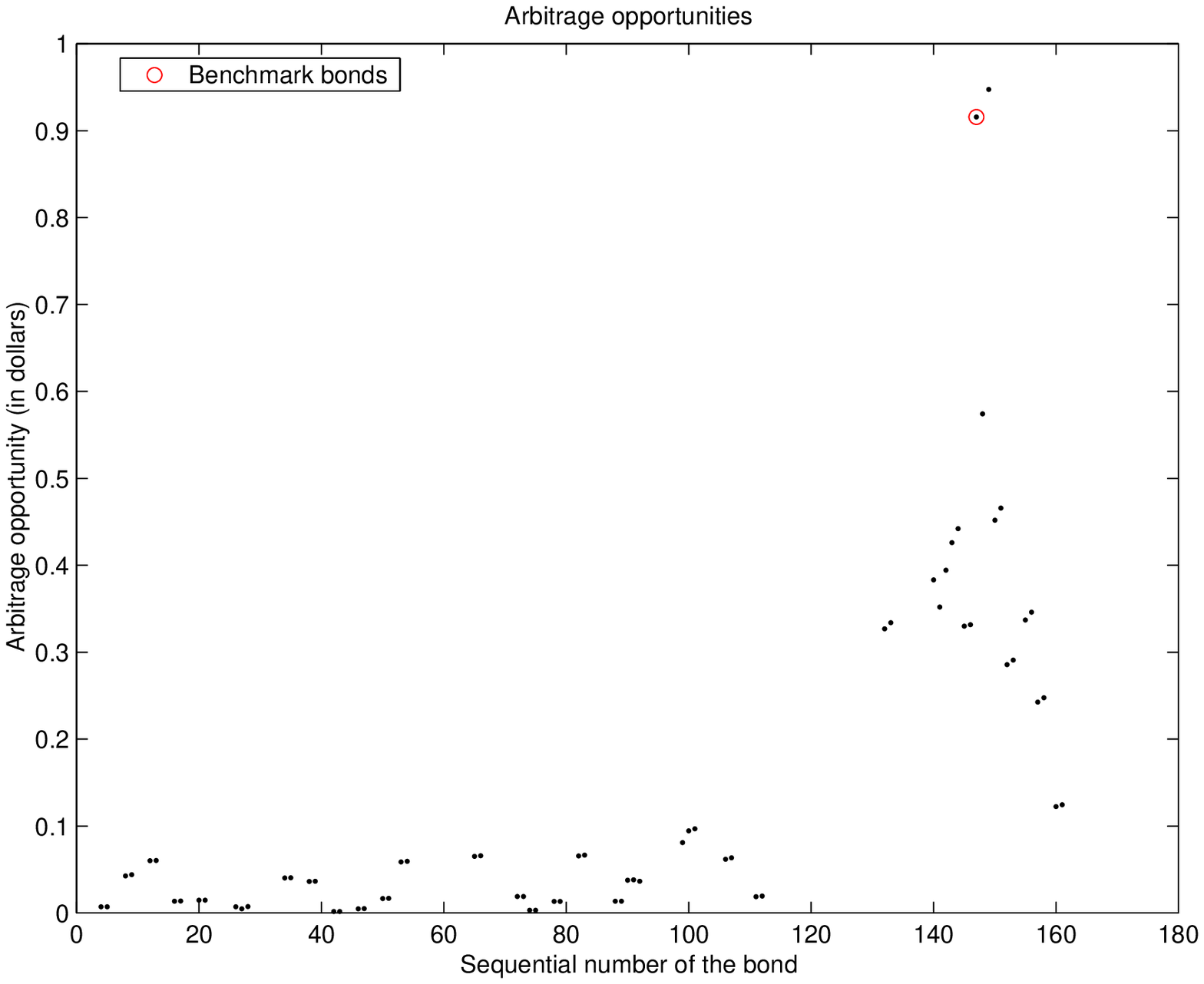} 
\end{center} 
\caption{\label{fig:arbitraggio}} 
\end{figure}

As first step, we built all minimal replicating portfolios (if any) of each
available bond to double check whether condition
\ref{cond:arbitraggio 1} is actually violated by bonds in the data set.  As
a ``measure'' of the arbitrage opportunities for a bond, we took the
largest absolute value of the difference between the price of the bond and
the price of all its minimal replicating portfolios.  The results of this
test are reported in the figure \ref{fig:arbitraggio}.
 
Note that just $69$ bonds, out of $206$, admit a (minimal) 
replicating portfolio, the maturity of such bonds reaching at most 
$7$ years. This is easily seen by comparing figures 
\ref{fig:arbitraggio} and \ref{fig:spettro}.  Moreover arbitrage 
opportunities amount to less than $\$ 0.1$ for bonds with maturity 
up to $3$ years, and range between $\$ 0.2$ and $\$ 1$ for bonds 
with maturity within the range $3-7$ years. 
 
To estimate the term structure of interest rates we have resorted 
to the approaches suggested by Carleton and Cooper 
\cite{CarletonCooper76}, to the spline approaches by McCulloch 
\cite{McCulloch71,McCulloch75} and by Coleman, Fisher and Ibbotson 
\cite{ColemanFisherIbbotson92} and to the parsimonious model by 
Nelson and Siegel \cite{NelsonSiegel87} (see section 
\ref{sect:stima della term structure} for a description of 
these methods). The technique proposed by Svensson 
\cite{Svensson94} gave us the same kind of results as that
proposed by Nelson and Siegel. Hence this set of experiments will 
not be reported in the present paper. 
 
To apply the method of Carleton and Cooper to a bond sample, $\S$, 
whose maturity spectrum contains "gaps", it is necessary
to shrink the sample itself to a suitable subset, $\T$, that 
forms a complete coupon term structure. After that we can estimate 
the discount factors by a constrained least squares procedure as 
described in section \ref{sect:stima della term structure}. In 
our case the subset $\T$ contains $152$ (out of $173$) US 
Treasury notes and bonds (the longest of which has maturity of 
about $7$ years) and, obviously, all of $33$ US Treasury bills. 
 
\begin{figure} 
\begin{center} 
\includegraphics*[scale=0.40]{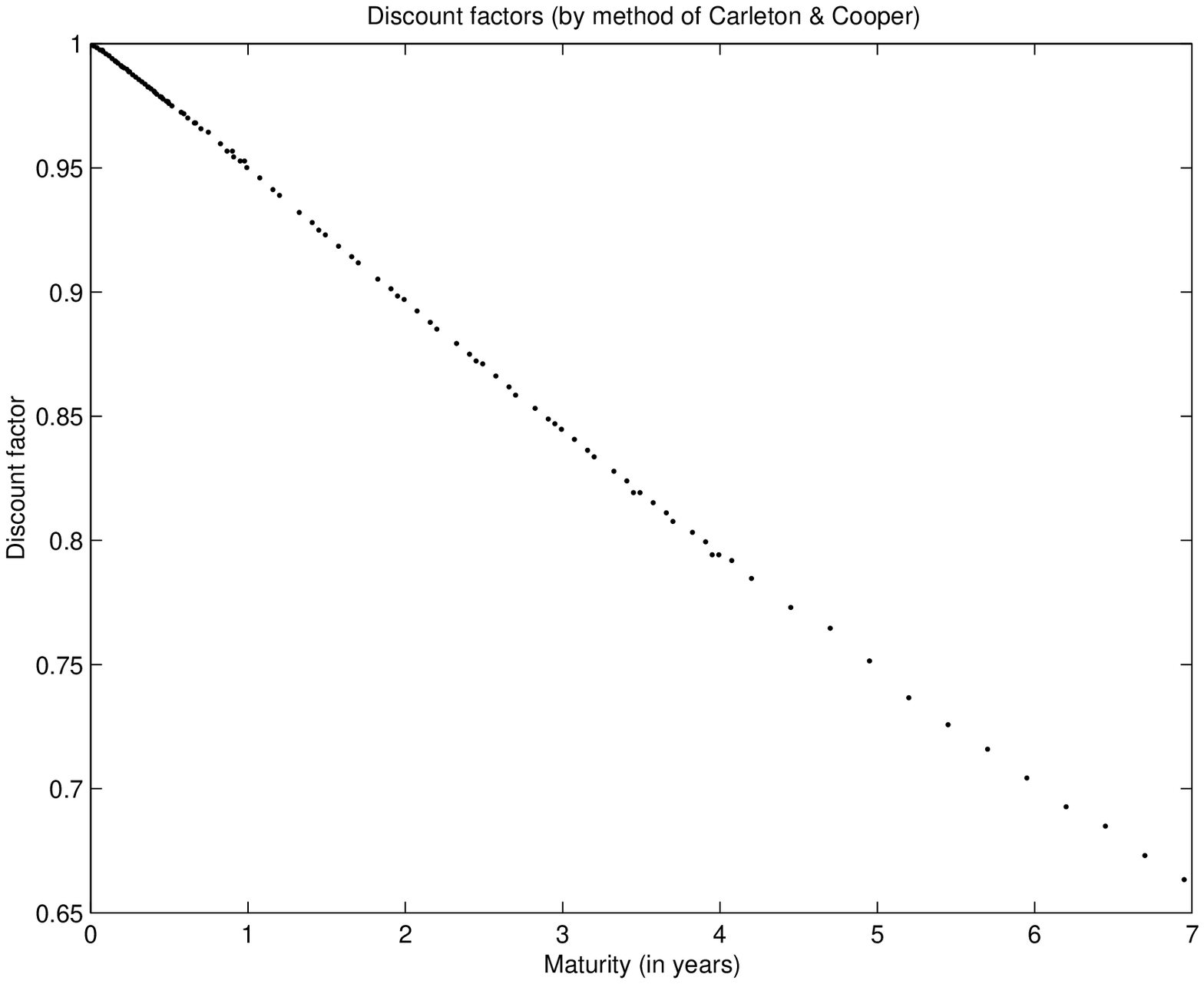} 
\includegraphics*[scale=0.40]{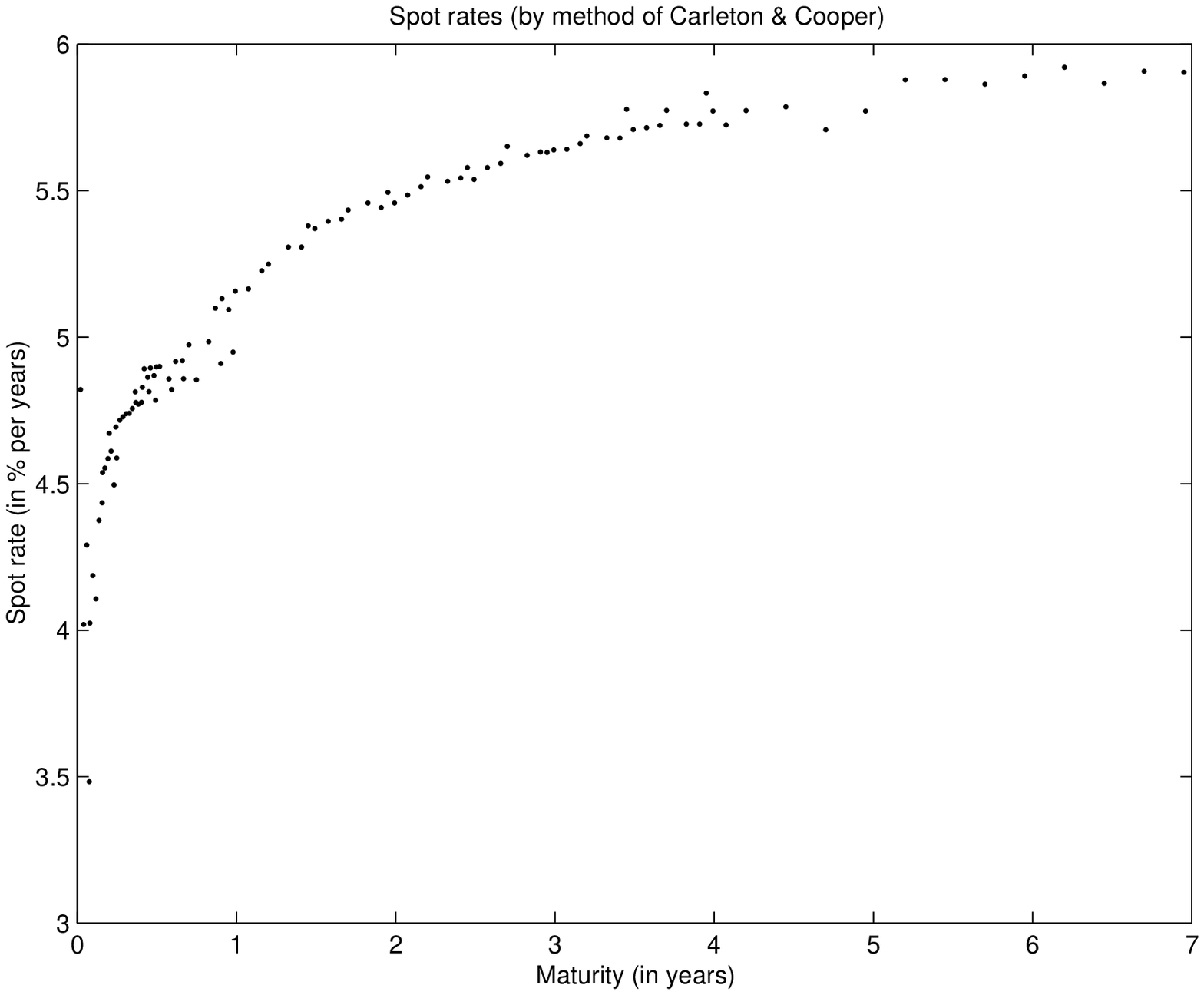} 
\includegraphics*[scale=0.40]{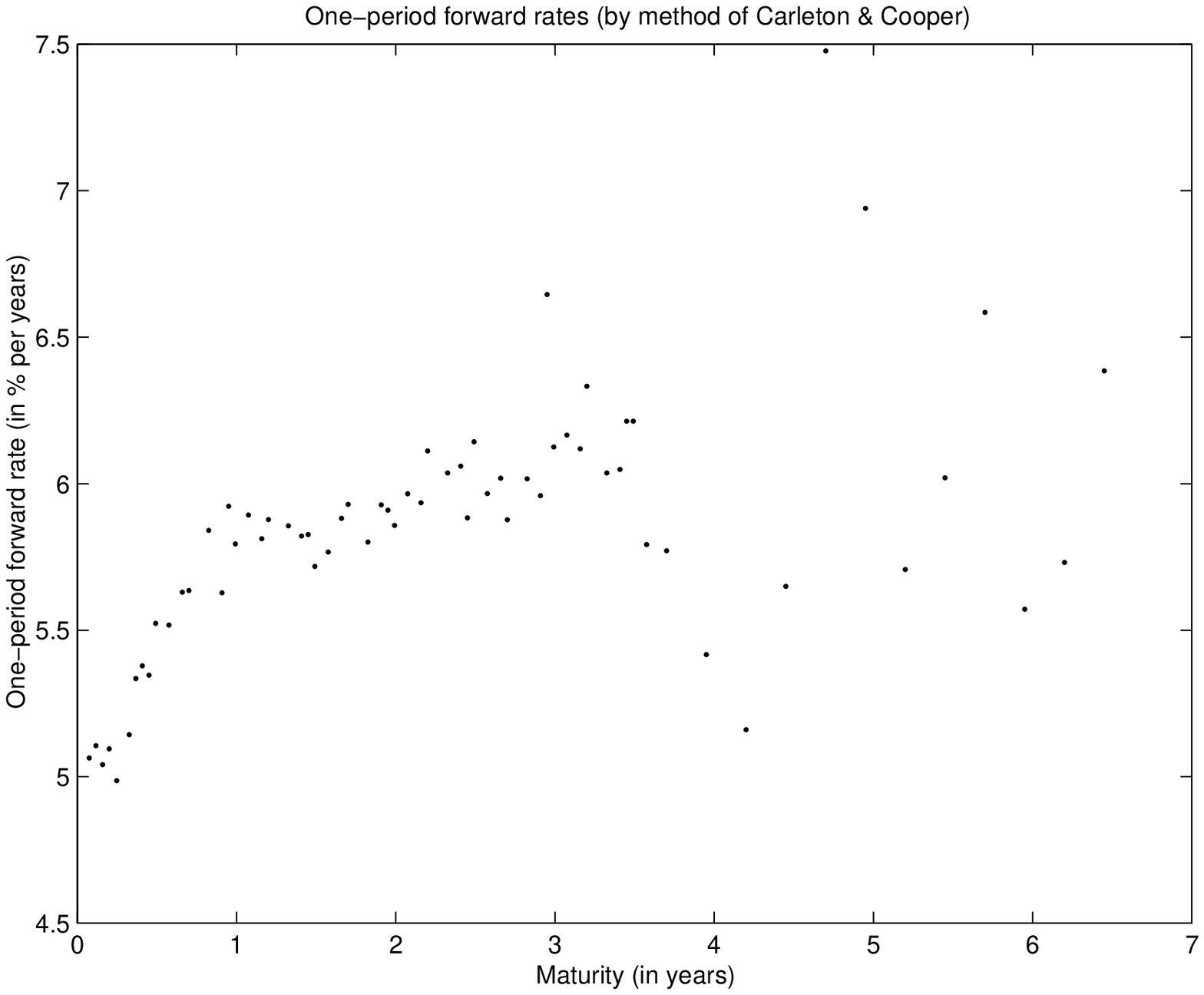} 
\includegraphics*[scale=0.40]{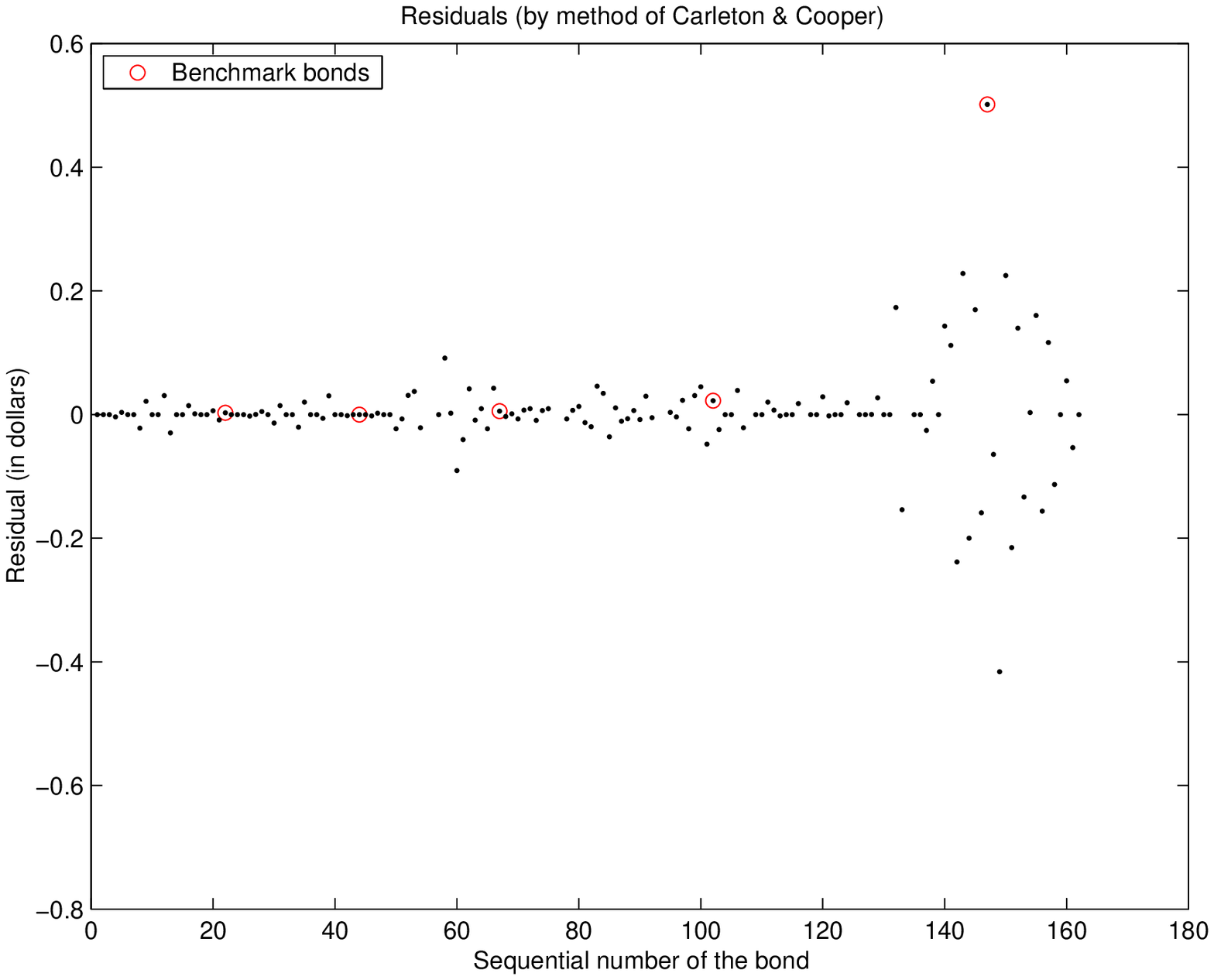} 
\end{center} 
\caption{\label{fig:Carleton}} 
\end{figure}

For our numerical experiments, neither the covariance matrix, 
$\sigma_{ij}=E\{\varepsilon_i \varepsilon_j\}$ nor other 
information about the statistical properties of the errors
$\varepsilon_j$'s was available. For such reason we have resorted 
to a straight least squares procedure, instead of a weighted one, 
so as to obtain unbiased, even though not minimum variance, 
estimators of the discount factors. Our results are reported in 
figure \ref{fig:Carleton}. 
 
Note that the discount factors approach one as their maturity 
tends to zero. The discount factors vary quite regularly with maturity, 
unlike spot rates and, even more significantly, one-period forward 
rates. These present irregularities for maturities of about $1$ 
year and above $3$ years, where the largest violations of 
condition \ref{cond:arbitraggio 1} occur. The plot of the 
residuals 
\begin{equation*} 
    p_j-\sum_{i=1}^N \hat{d}_i\varphi_i(c_j,m_j), 
\end{equation*} 
shows that, as expected, the fitting procedure performs worse 
where condition \ref{cond:arbitraggio 1} is more severely 
violated. 
 
\begin{figure} 
\begin{center} 
\includegraphics*[scale=0.40]{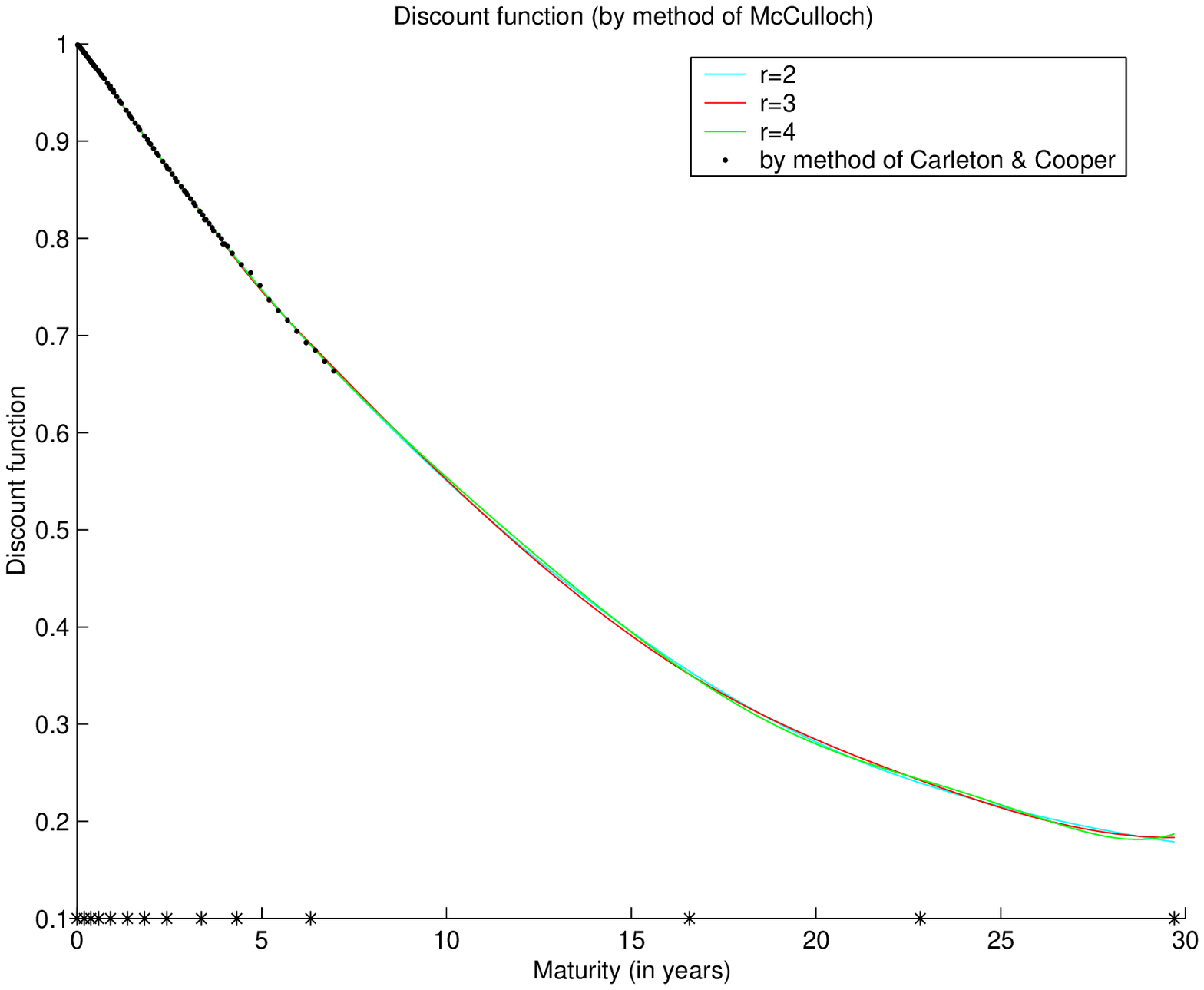} 
\includegraphics*[scale=0.40]{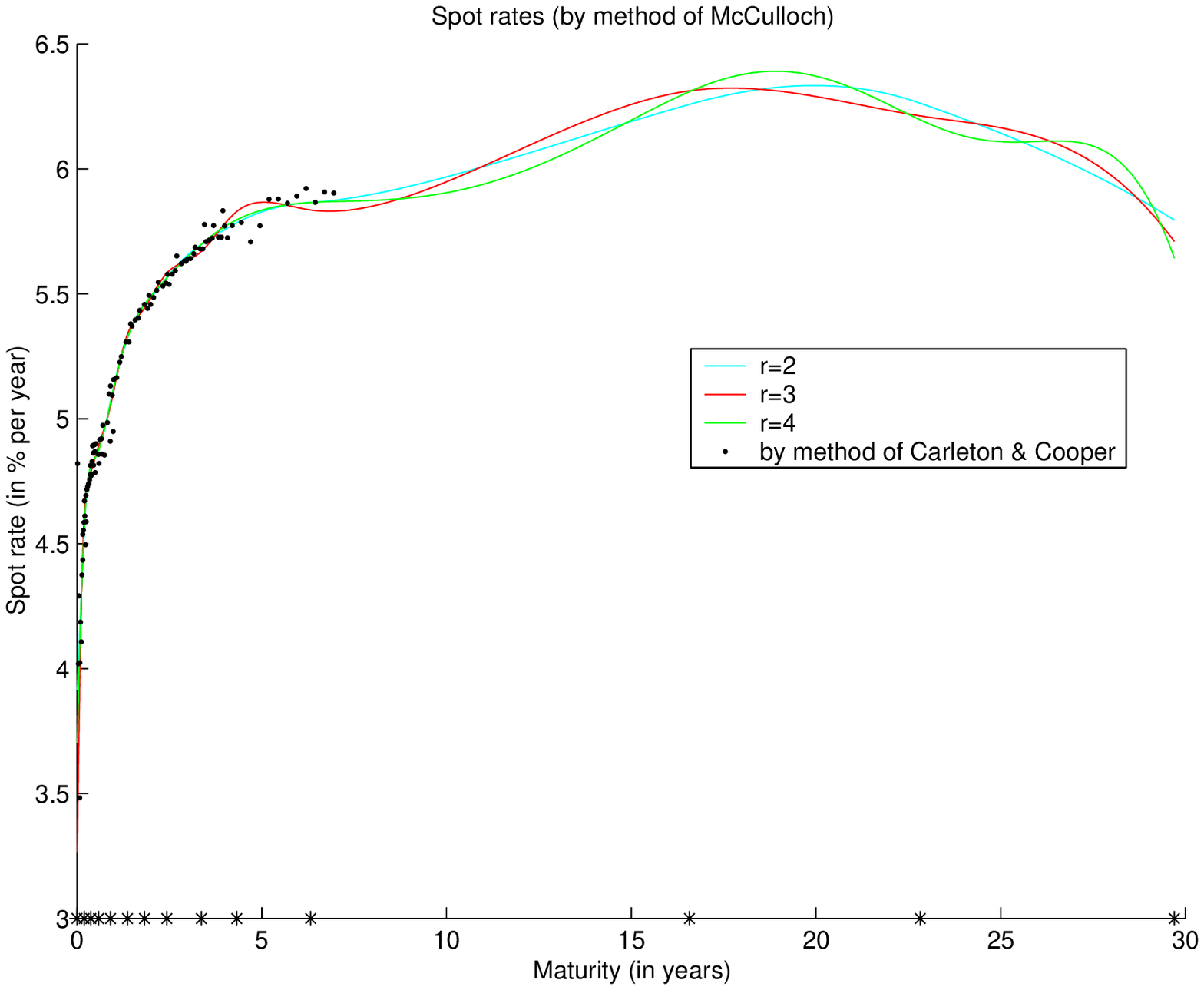} 
\includegraphics*[scale=0.40]{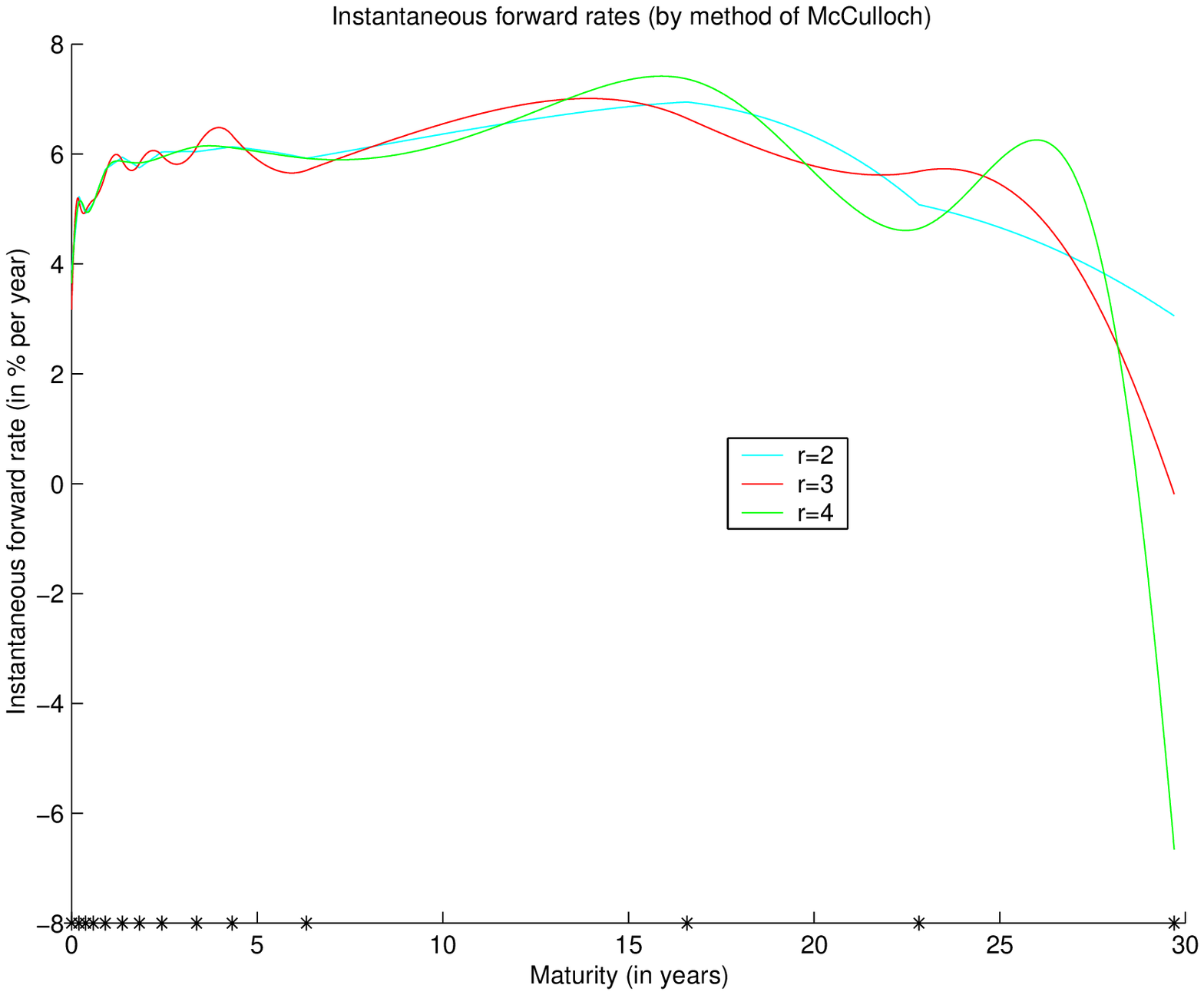} 
\includegraphics*[scale=0.40]{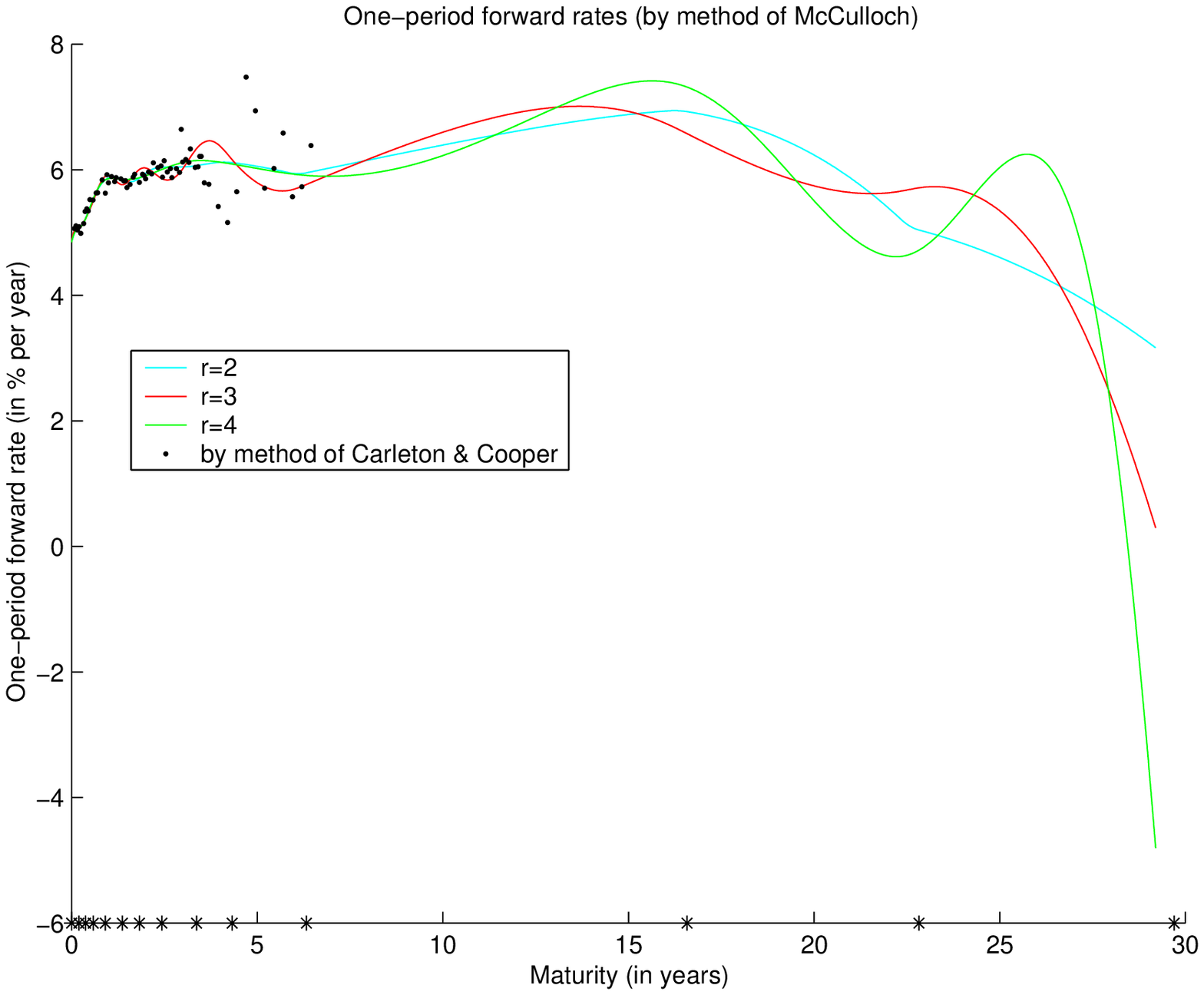} 
\end{center} 
\caption{\label{fig:McCulloch_McCulloch}} 
\end{figure}

\begin{figure} 
\begin{center} 
\includegraphics*[scale=0.40]{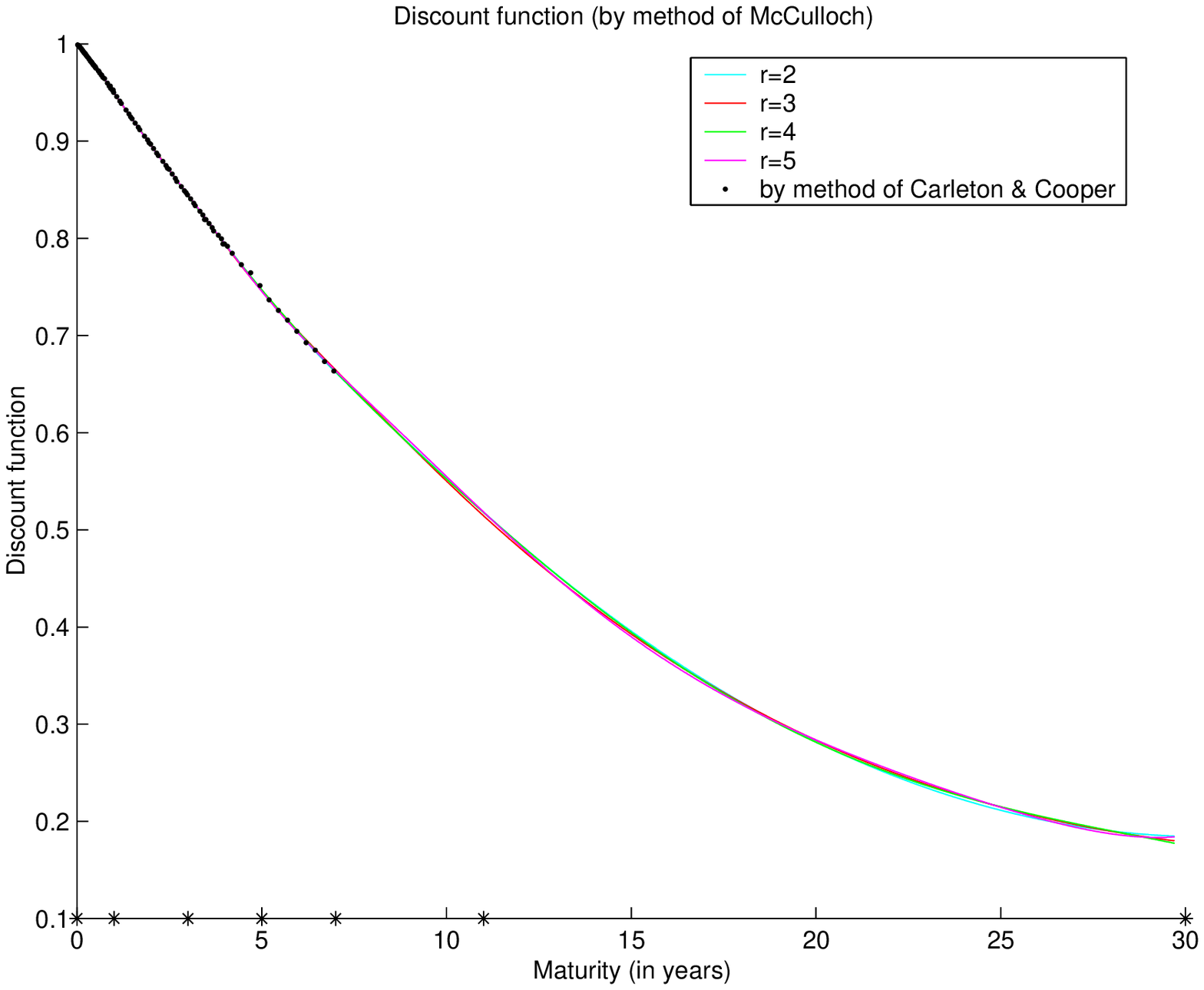} 
\includegraphics*[scale=0.40]{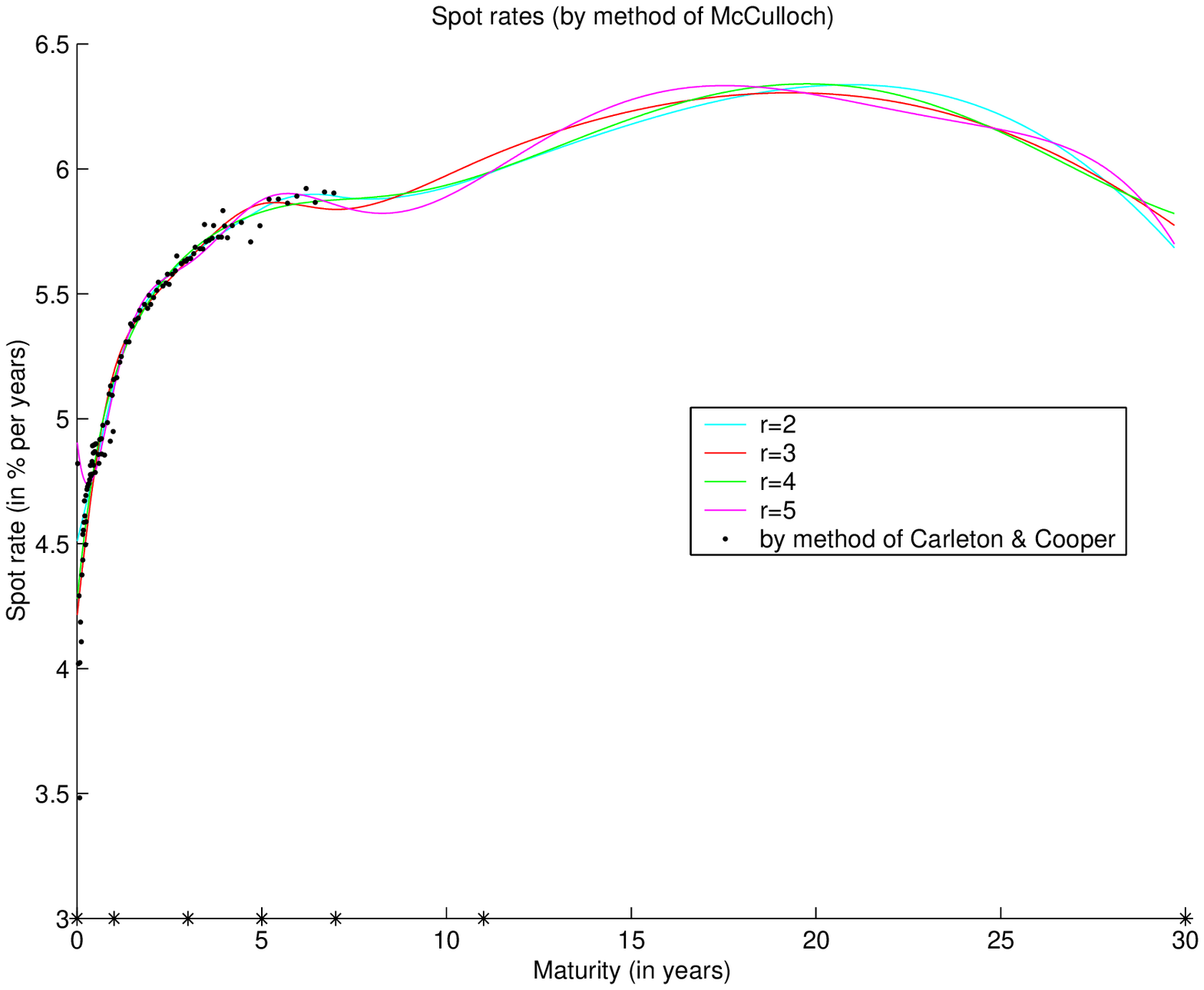} 
\includegraphics*[scale=0.40]{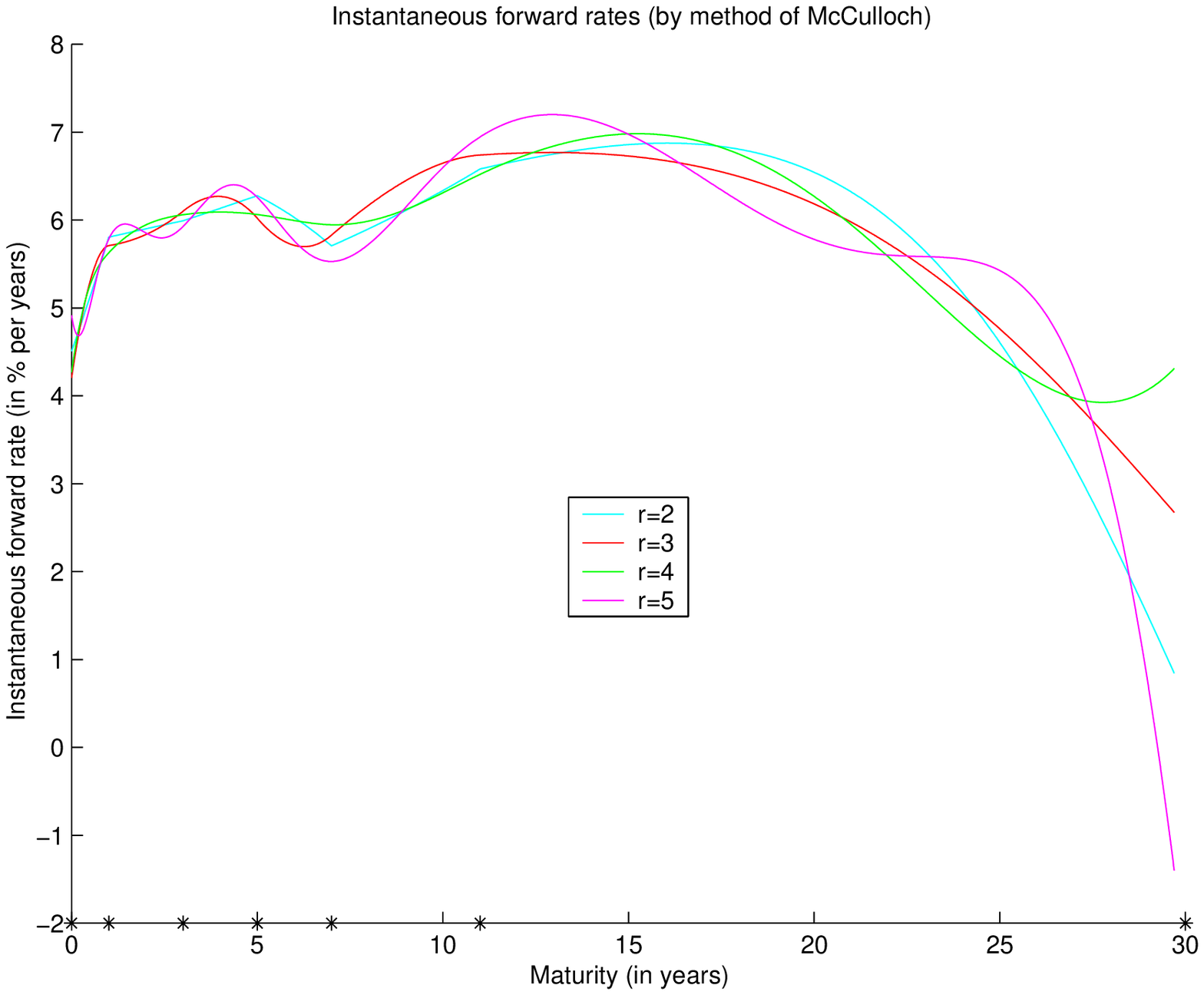} 
\includegraphics*[scale=0.40]{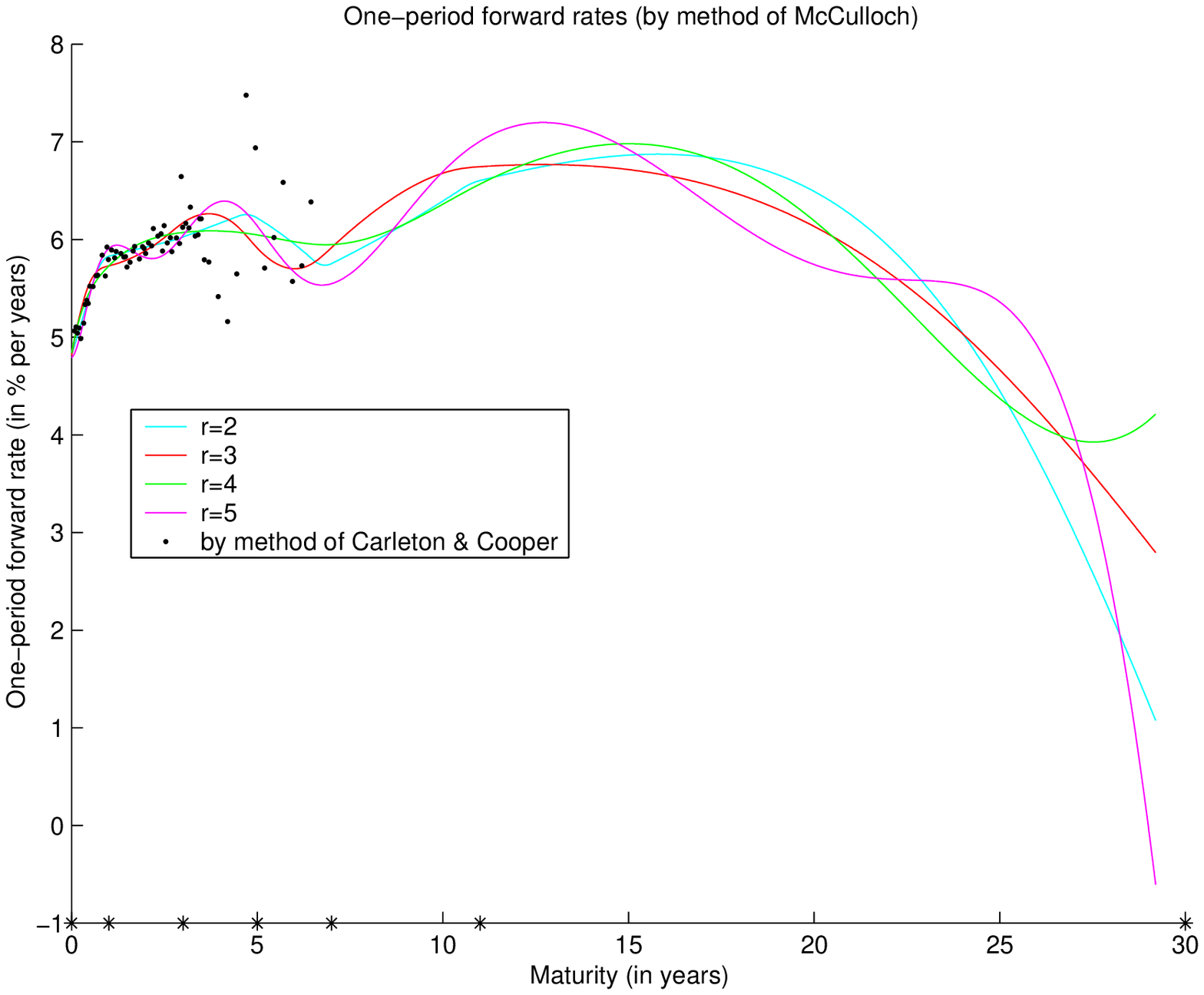} 
\end{center} 
\caption{\label{fig:McCulloch_1-3-5-7-11}} 
\end{figure}

McCulloch's method has been tested on our data sample with two
different sets of knot points.  The first set (hereafter referred to
as ``knots {\it ala} McCulloch'') is built following McCulloch's
indication and consists of $14$ knots placed at $0$, $0.2$, $0.4$,
$0.6$, $0.9$, $1.4$, $1.8$, $2.4$, $3.4$, $4.3$, $6.3$, $16.6$,
$22.8$, $29.7$ years.  The second set is made of seven knots placed at
$0$, $1$, $3$, $5$, $7$, $11$ and $30$ years.  The base we used (see
section \ref{subsec:metodi spline}) consists of the following
functions:
\begin{equation*}
    \begin{align*} 
        f_i(t) &=t^i \qquad (i=0,\ldots,r-1), \\ 
        f_i(t) &= 
        \begin{cases} 
	    0 &, \quad\text{if}\; t\leq K_{i-r+1}, \\
	    (t-K_{i-r+1})^r &, \quad\text{if}\; K_{i-r+1}<t\leq
K_{i-r+2}, \\
	    (t-K_{i-r+1})^r -(t-K_{i-r+2})^r &, \quad\text{if}\;
K_{i-r+2}<t
        \end{cases} \\
        & \hspace{7cm} (i=r,\ldots,r+k-2).  
    \end{align*}
\end{equation*}
and polynomial spline functions up to the fifth degree have been used. 
Our results are reported in figure \ref{fig:McCulloch_McCulloch} 
for the first set of knots, and in figure 
\ref{fig:McCulloch_1-3-5-7-11} for the second set. 
 
These graphs show all drawbacks of McCulloch's method.  First the
outcome of the regression procedure depends strongly on the positions
of the knots and on the degree of the polynomial spline employed.
This feature is apparent in the forward rate curves (compare figure
\ref{fig:McCulloch_McCulloch} and \ref{fig:McCulloch_1-3-5-7-11})
whereas the fits for the discount function and the spot rate are more
stable.  Moreover, a polynomial spline is not, in general, a
decreasing function, hence negative forward rates may occur.  It is
interesting to note how results for one-period forward rates
obtained both by the method of McCulloch (or Coleman, Fisher and
Ibbotson as we will see soon) and by the method of Carleton and Cooper
share the same oscillating behaviour within $3$ and $7$ years. This
suggests that such behaviour actually depends on the violations
of the condition of absence of arbitrage opportunities that occur at
those maturities (see figure \ref{fig:arbitraggio}).

To assess the criticism of Vasicek and Fong \cite{VasicekFong82} 
to McCulloch's technique (see section \ref{sect:stima della term 
structure}), we assumed an exponential discount function of the 
form 
\begin{equation}\label{eq:discount fittizia} 
    d(t)\equiv e^{-0.06 t}, 
\end{equation} 
which corresponds to a spot rate function and a forward rate 
function identically equal to $6\%$ per year, and generated two 
sets of artificial data as follows. For each bond 
$(c_j,m_j,p_j)$ in the original data set, $\S$, we defined a fake 
price: 
\begin{equation*} 
    \hat{p}_j :=\sum_{i=1}^N d(t_i)\varphi_i(c_j,m_j)+\varepsilon_j, 
\end{equation*} 
where $T(\S)=(t_1,\ldots,t_N)$ and $\varepsilon_j\equiv 0$ for the 
first set of artificial data (named hereafter ``exact'' data) 
whereas $\varepsilon_j\sim\mathcal{N}(0,1)$ for the second set 
(``noisy'' data). 
 
\begin{figure} 
\begin{center} 
\includegraphics*[scale=0.40]{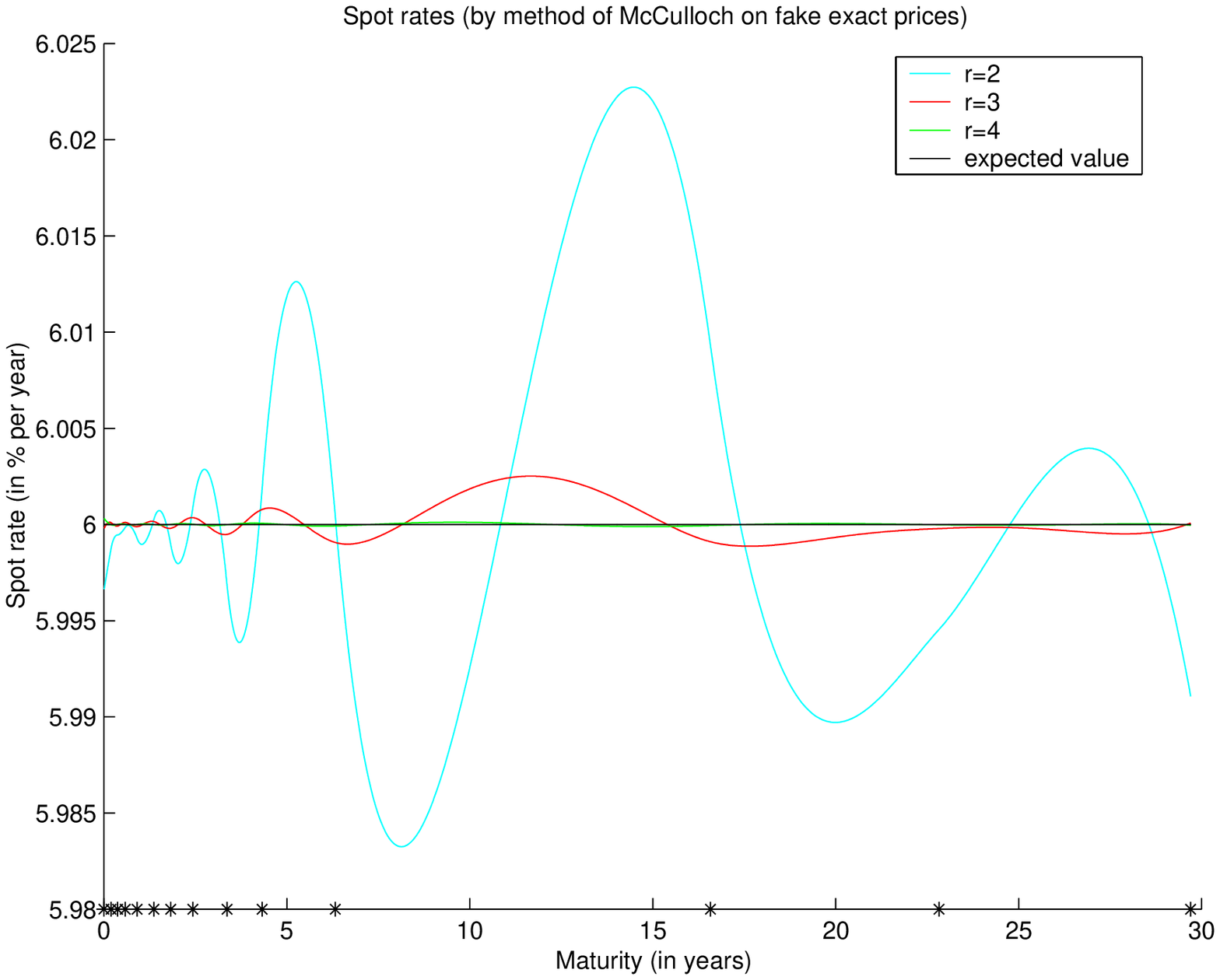} 
\includegraphics*[scale=0.40]{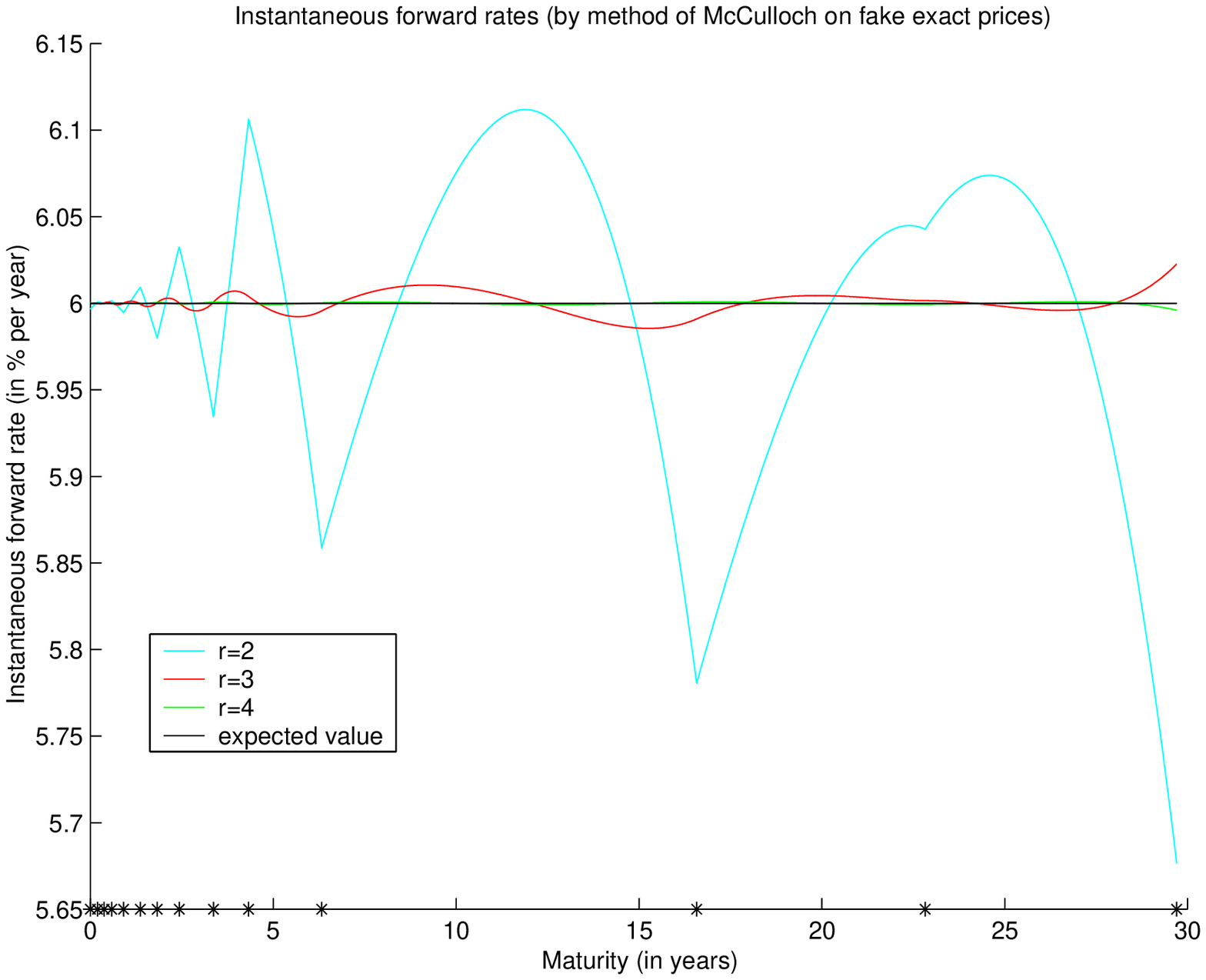} 
\end{center} 
\caption{\label{fig:esperimento_esatto}} 
\end{figure}

\begin{figure} 
\begin{center} 
\includegraphics*[scale=0.40]{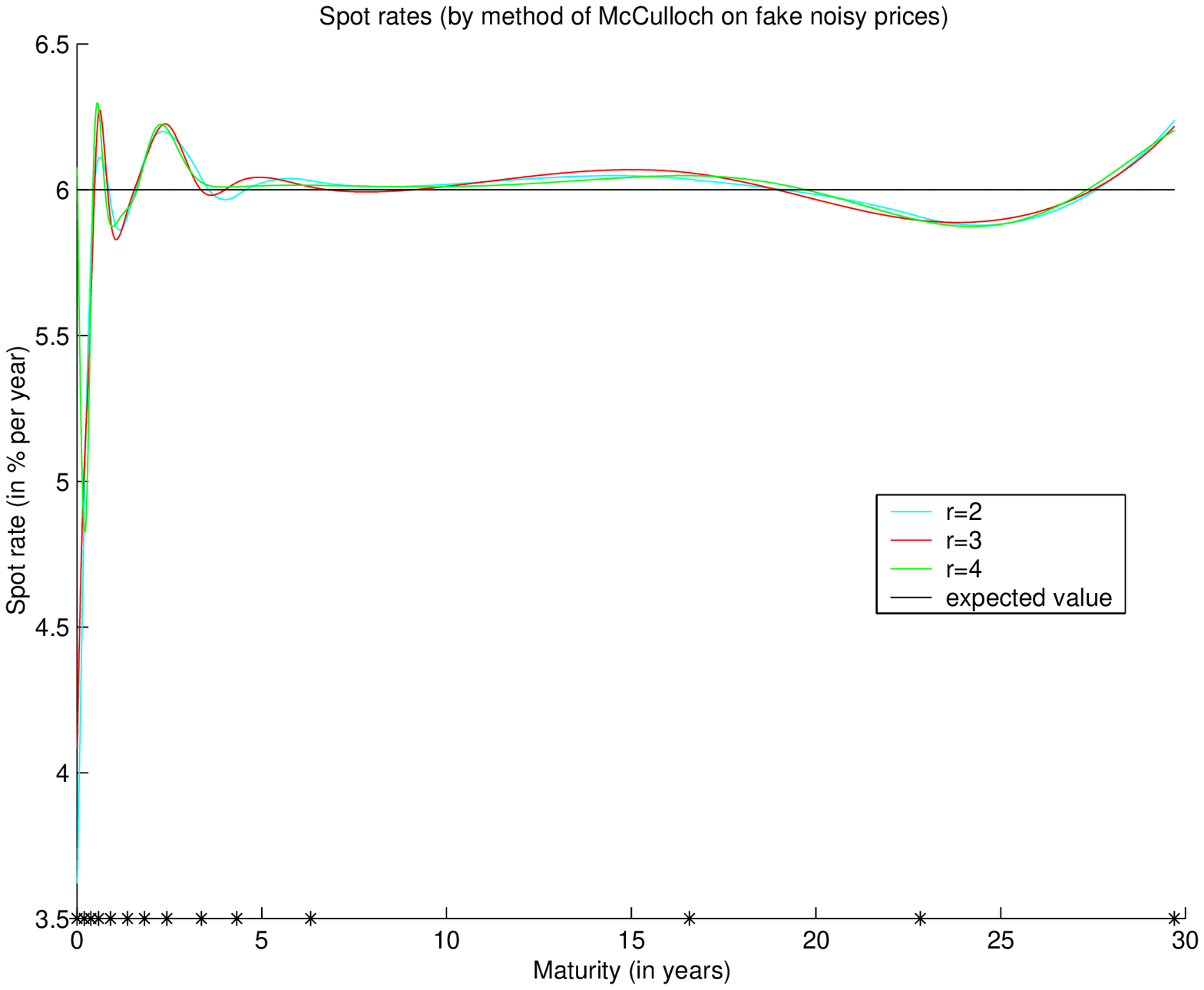} 
\includegraphics*[scale=0.40]{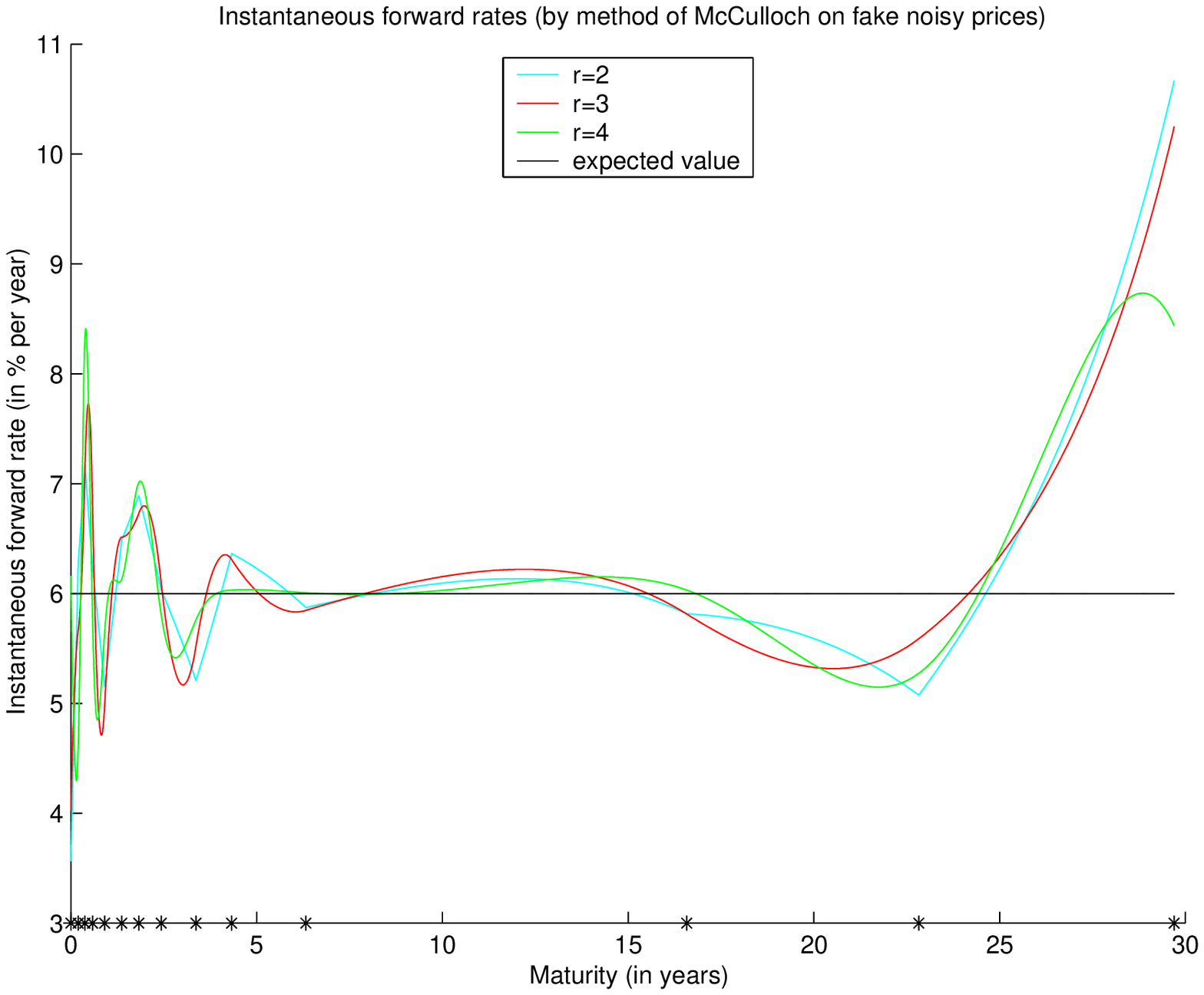} 
\end{center} 
\caption{\label{fig:esperimento_rumoroso}} 
\end{figure}

After that McCulloch's method, with knots {\it ala} McCulloch, has
been applied to both ``exact'' and ``noisy'' bond data so attaining an
estimate, $\hat{d}$, of the discount function $d$ defined in
\eqref{eq:discount fittizia}.  The results reported in figures
\ref{fig:esperimento_esatto} and \ref{fig:esperimento_rumoroso} seem
to confirm Vasicek and Fong's opinion.  McCulloch's method performs
pretty well on the ``exact'' data, whereas it provides a bad fit when
``noisy'' data are used, even if the artificial perturbations
introduced are quite small compared to the typical price of a bond
($<1\%$ on average).

\begin{figure} 
\begin{center} 
\includegraphics*[scale=0.40]{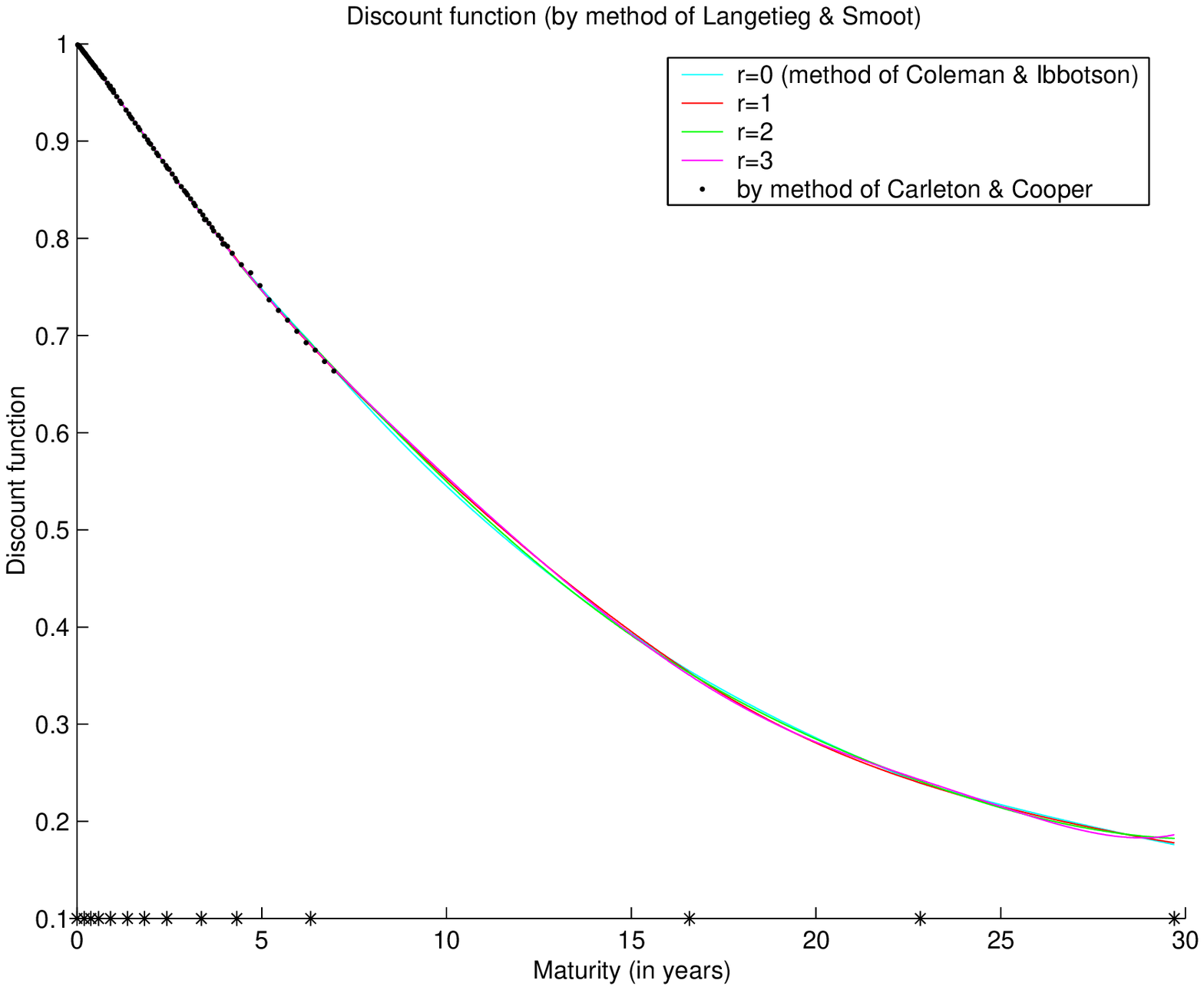} 
\includegraphics*[scale=0.40]{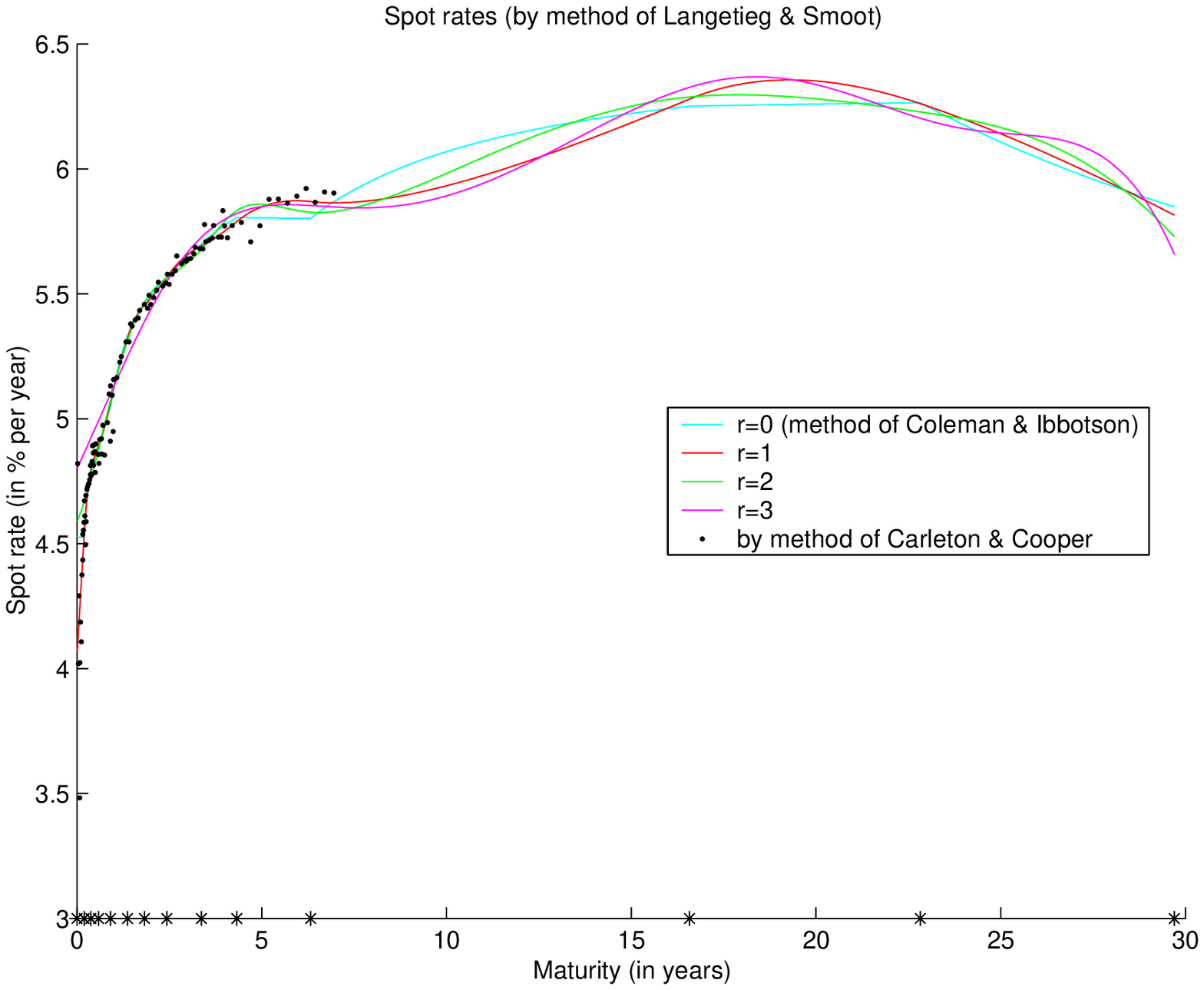} 
\includegraphics*[scale=0.40]{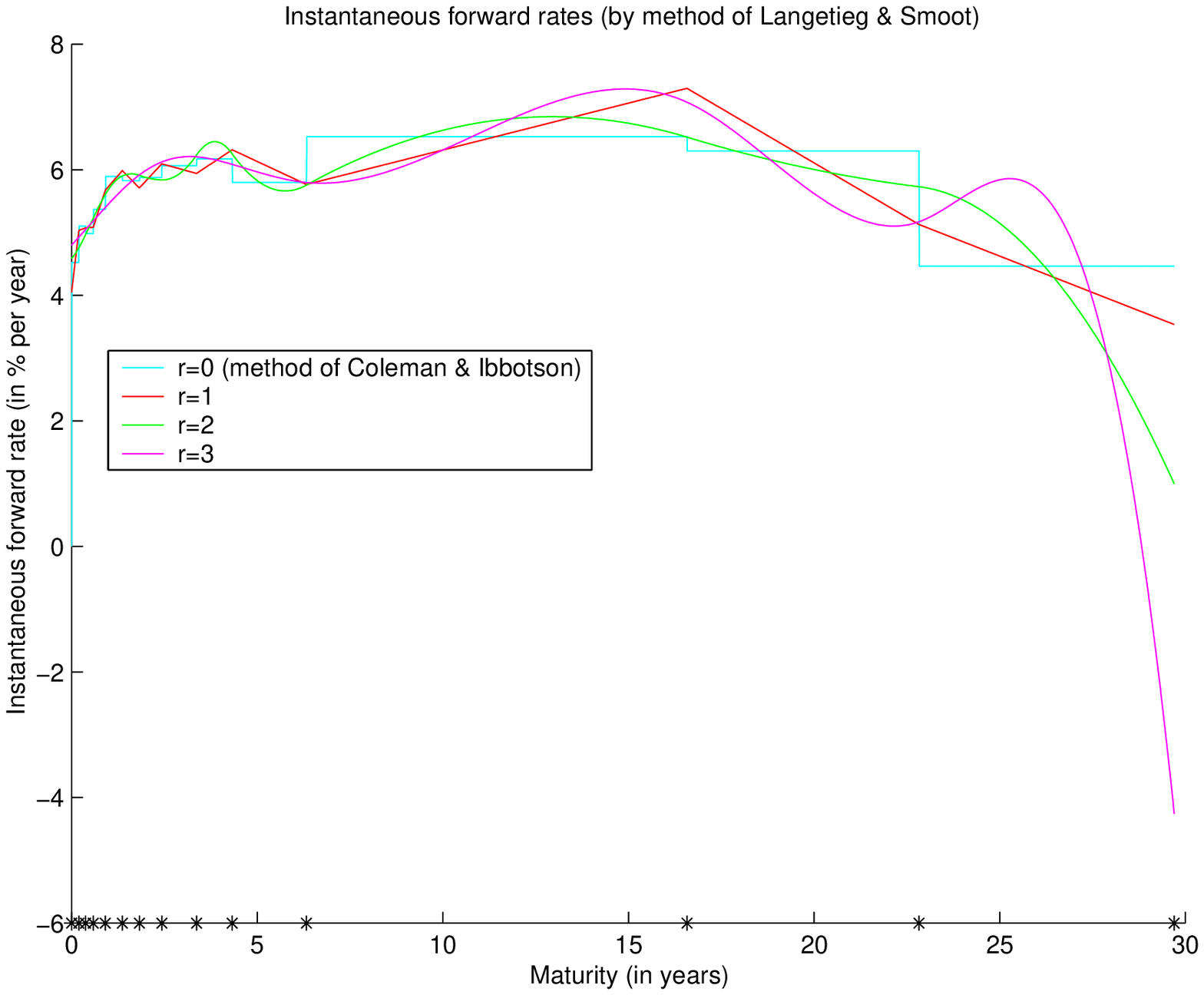} 
\includegraphics*[scale=0.40]{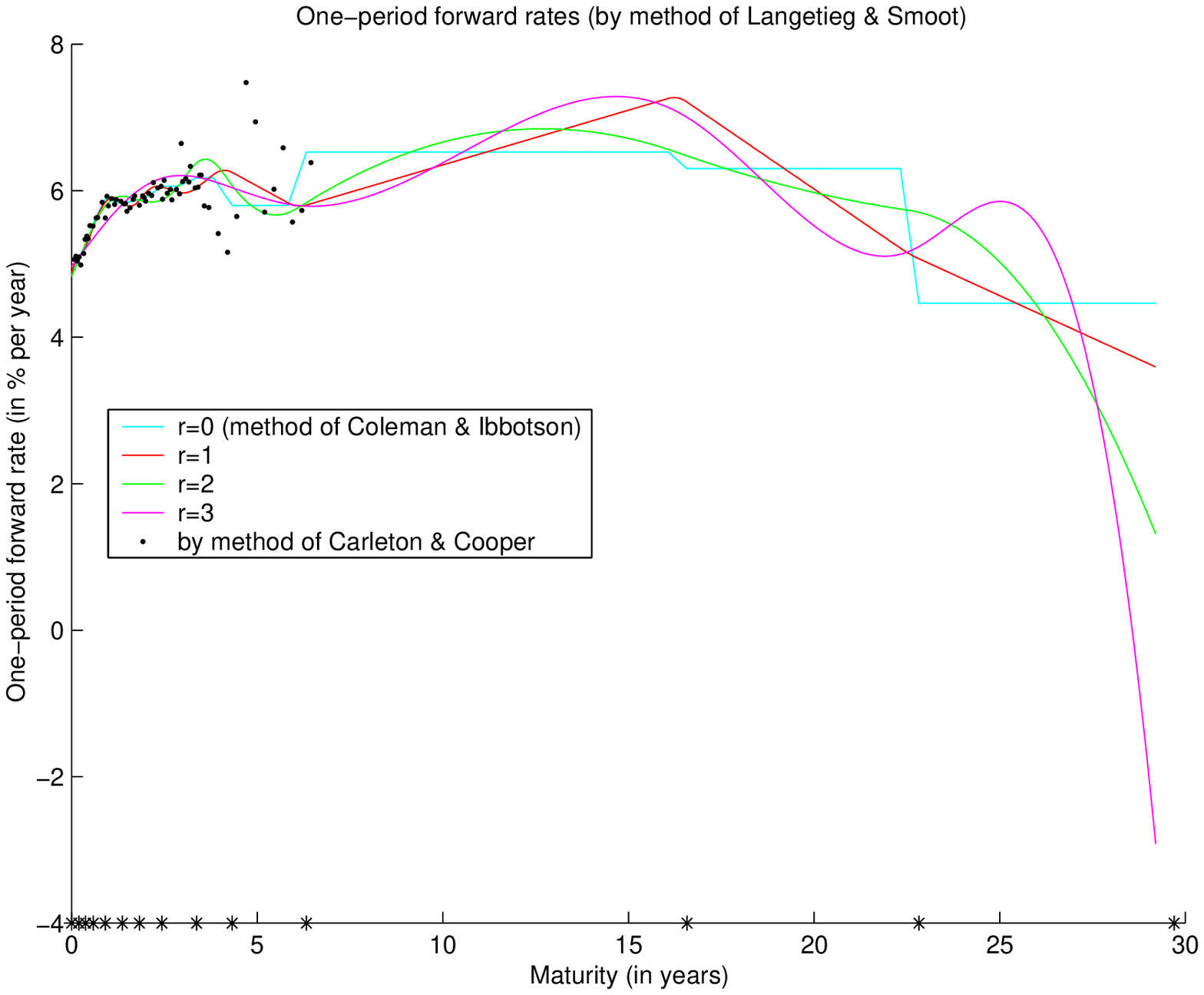} 
\end{center} 
\caption{\label{fig:Coleman_McCulloch}} 
\end{figure}

\begin{figure} 
\begin{center} 
\includegraphics*[scale=0.40]{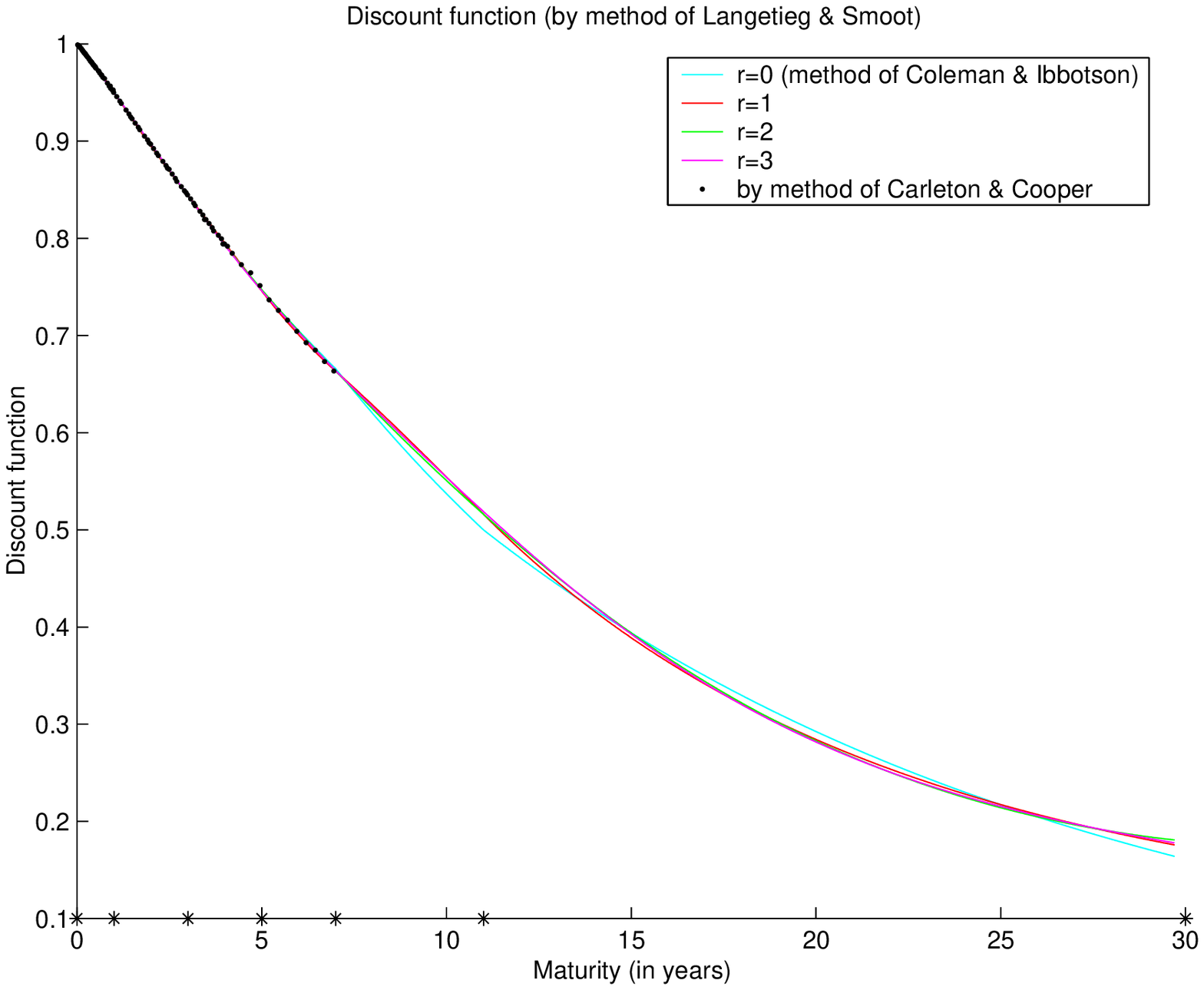} 
\includegraphics*[scale=0.40]{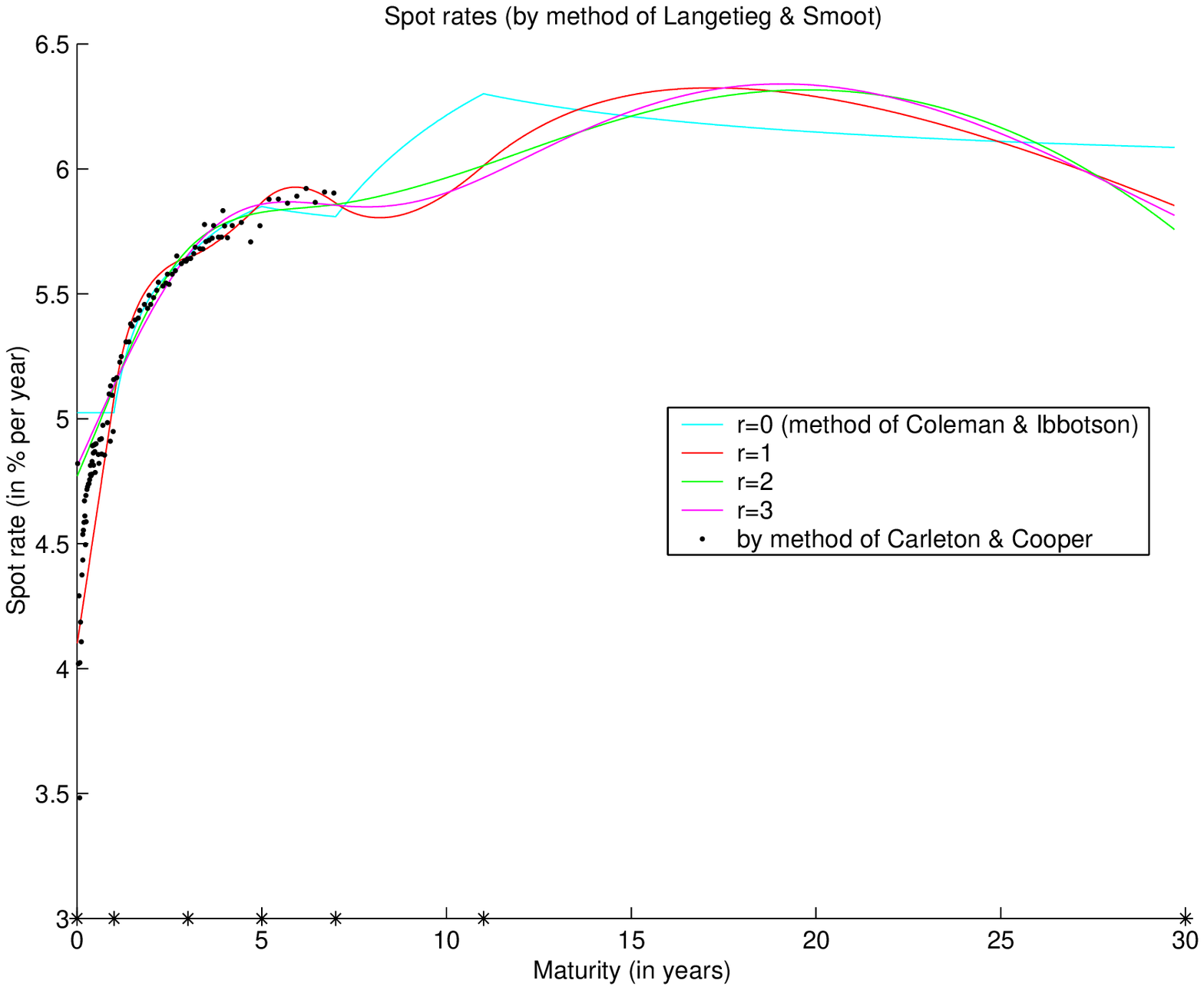} 
\includegraphics*[scale=0.40]{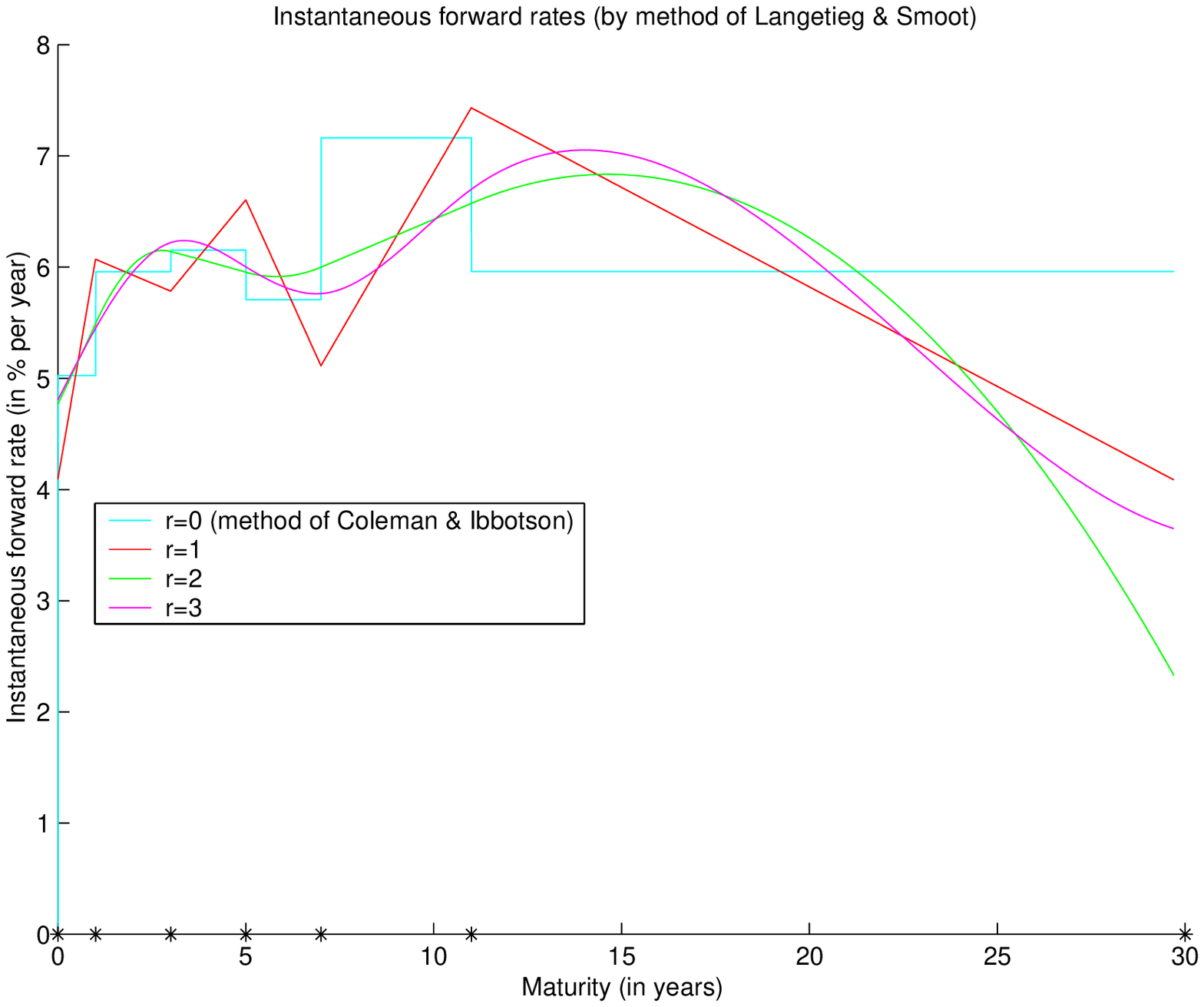} 
\includegraphics*[scale=0.40]{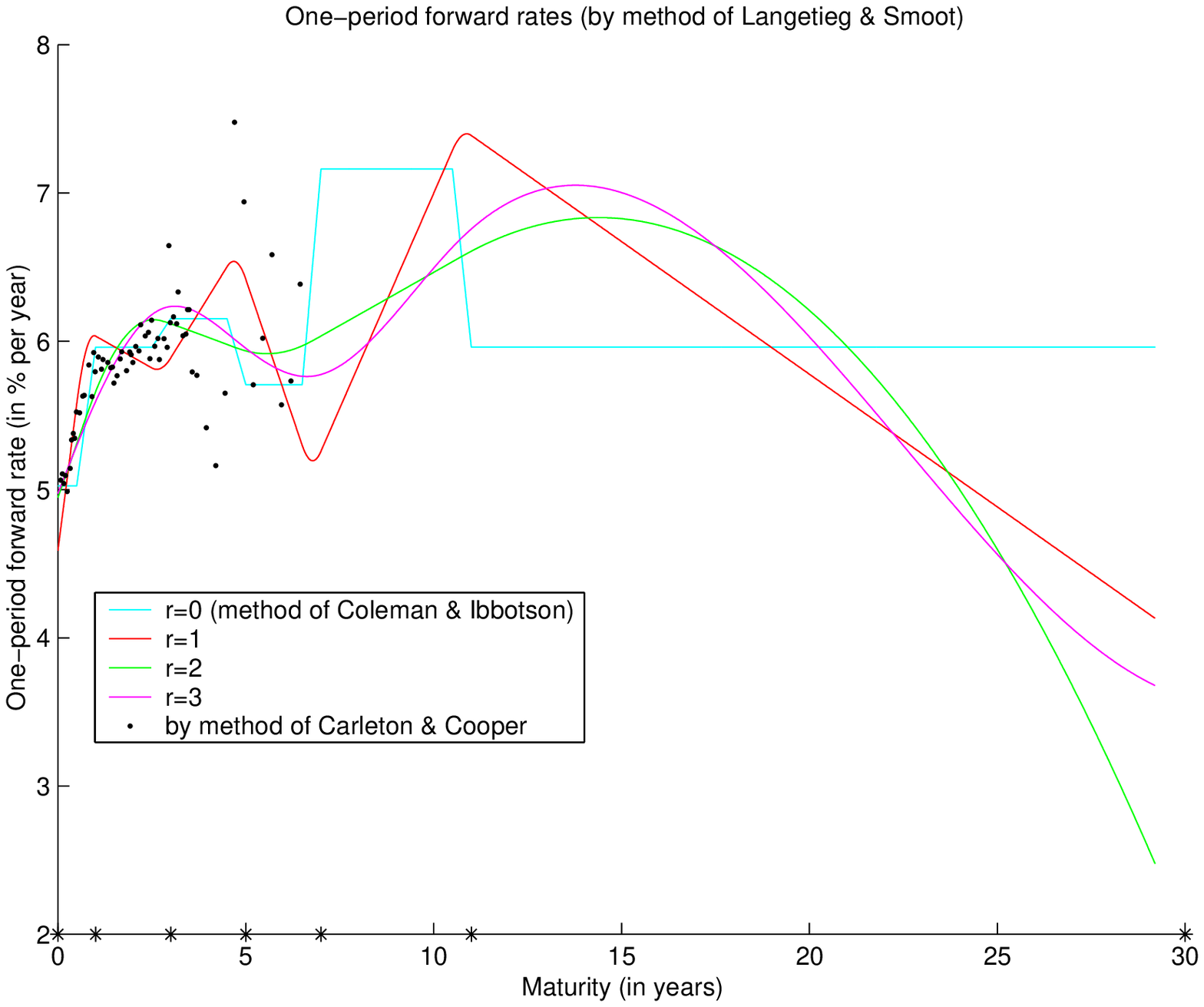} 
\end{center} 
\caption{\label{fig:Coleman_1-3-5-7-11}} 
\end{figure}

The method of Coleman, Fisher and Ibbotson has been tested with 
the same sets of knot points and the same base of spline functions 
that we used to test McCulloch's method. Polynomial spline 
functions of various degree have been used. For
$0$-degree spline functions, that is when a piecewise constant 
function is fitted to the instantaneous forward rate function, the 
least squares procedure has been constrained so as to attain a 
monotonic decreasing discount function. Our results are reported 
in figure \ref{fig:Coleman_McCulloch} for knots {\it ala} 
McCulloch, and in figure \ref{fig:Coleman_1-3-5-7-11} for knots 
placed at $0$, $1$, $3$, $5$, $7$, $11$ and $30$ years. 
 
These graphs show that the method of Coleman, Fisher and Ibbotson shares
the same drawbacks of McCulloch's method. However the risk of getting
negative forward rates seem to be reduced.
 
\begin{figure} 
\begin{center} 
\includegraphics*[scale=0.40]{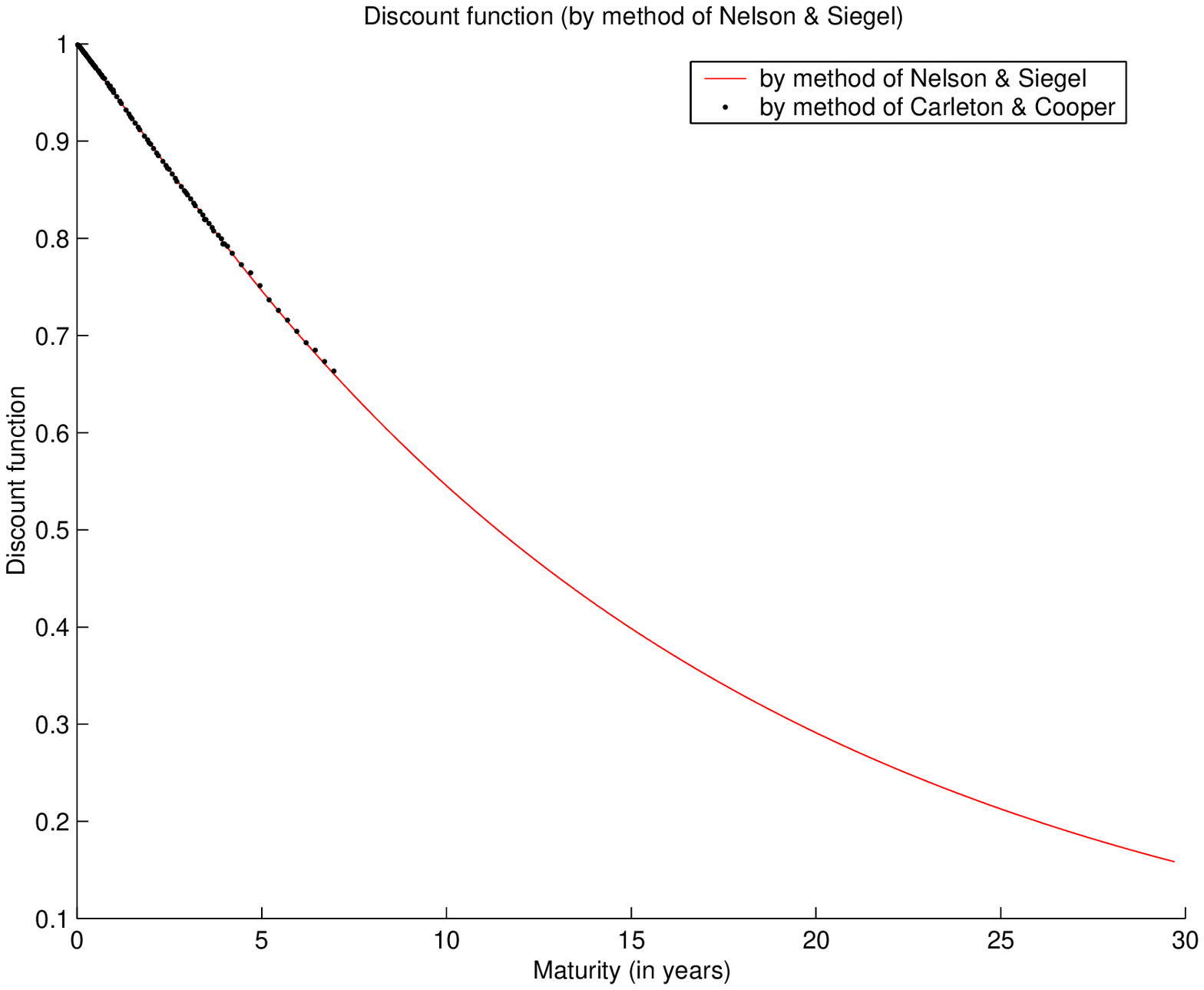} 
\includegraphics*[scale=0.40]{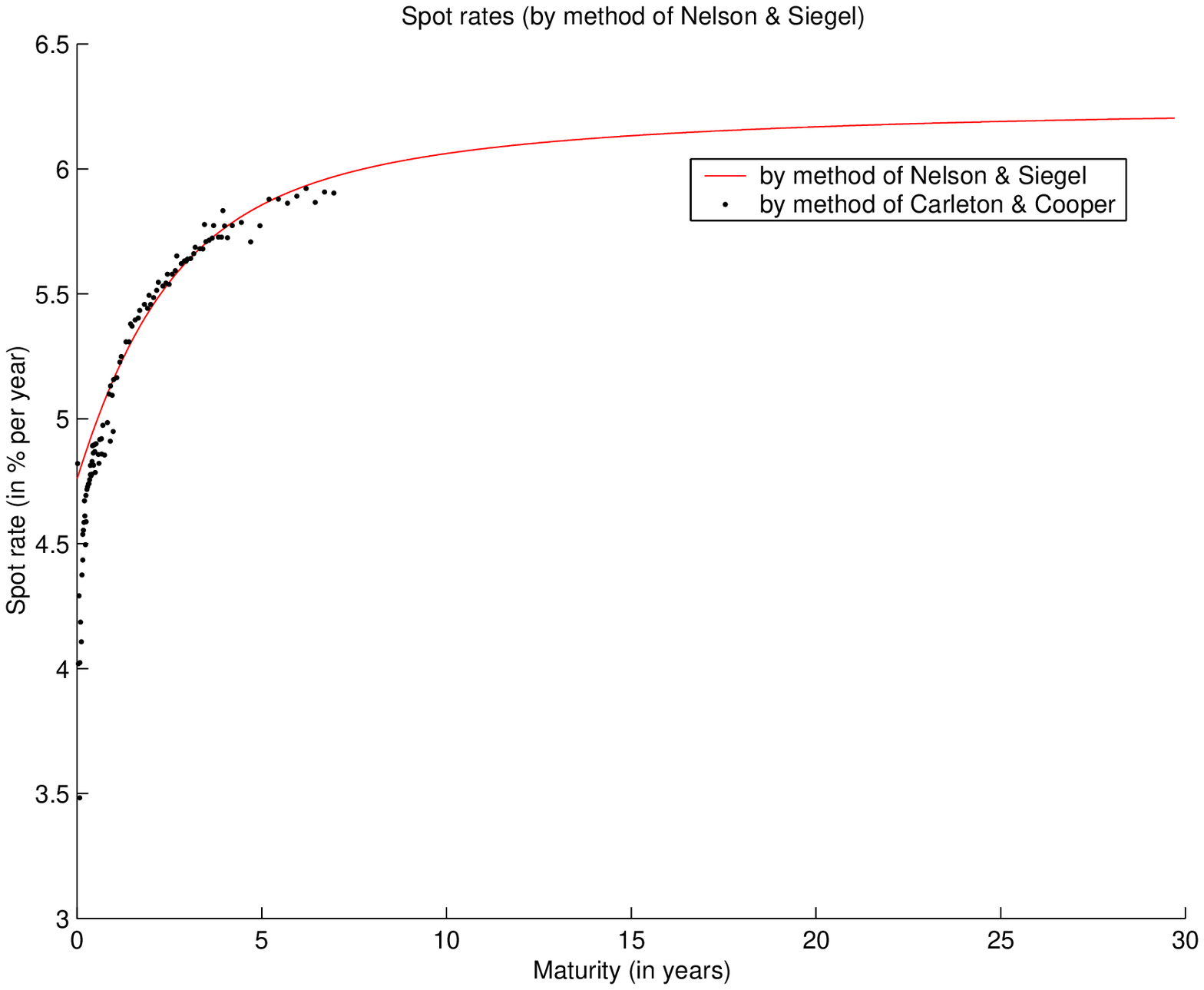} 
\includegraphics*[scale=0.40]{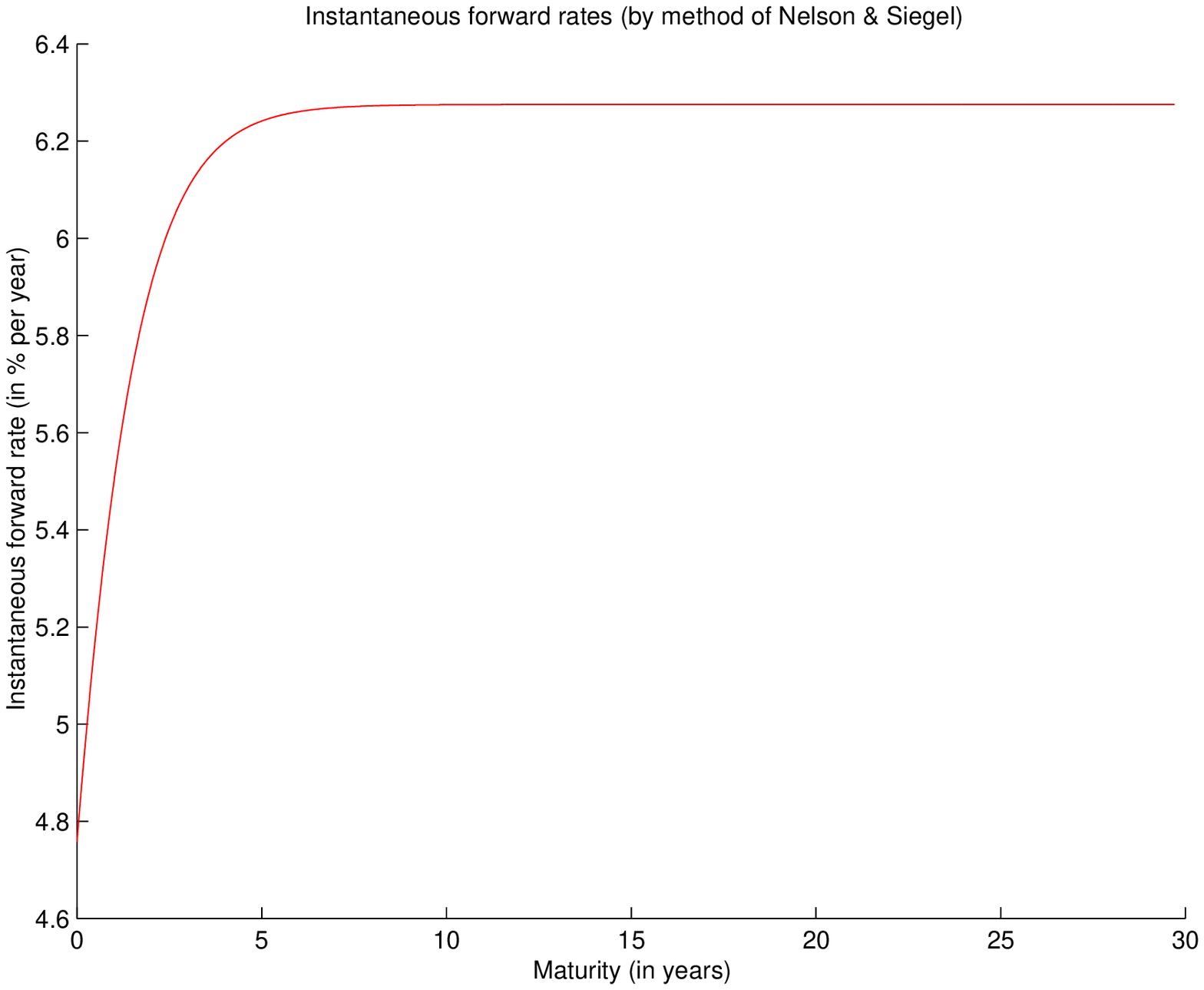} 
\includegraphics*[scale=0.40]{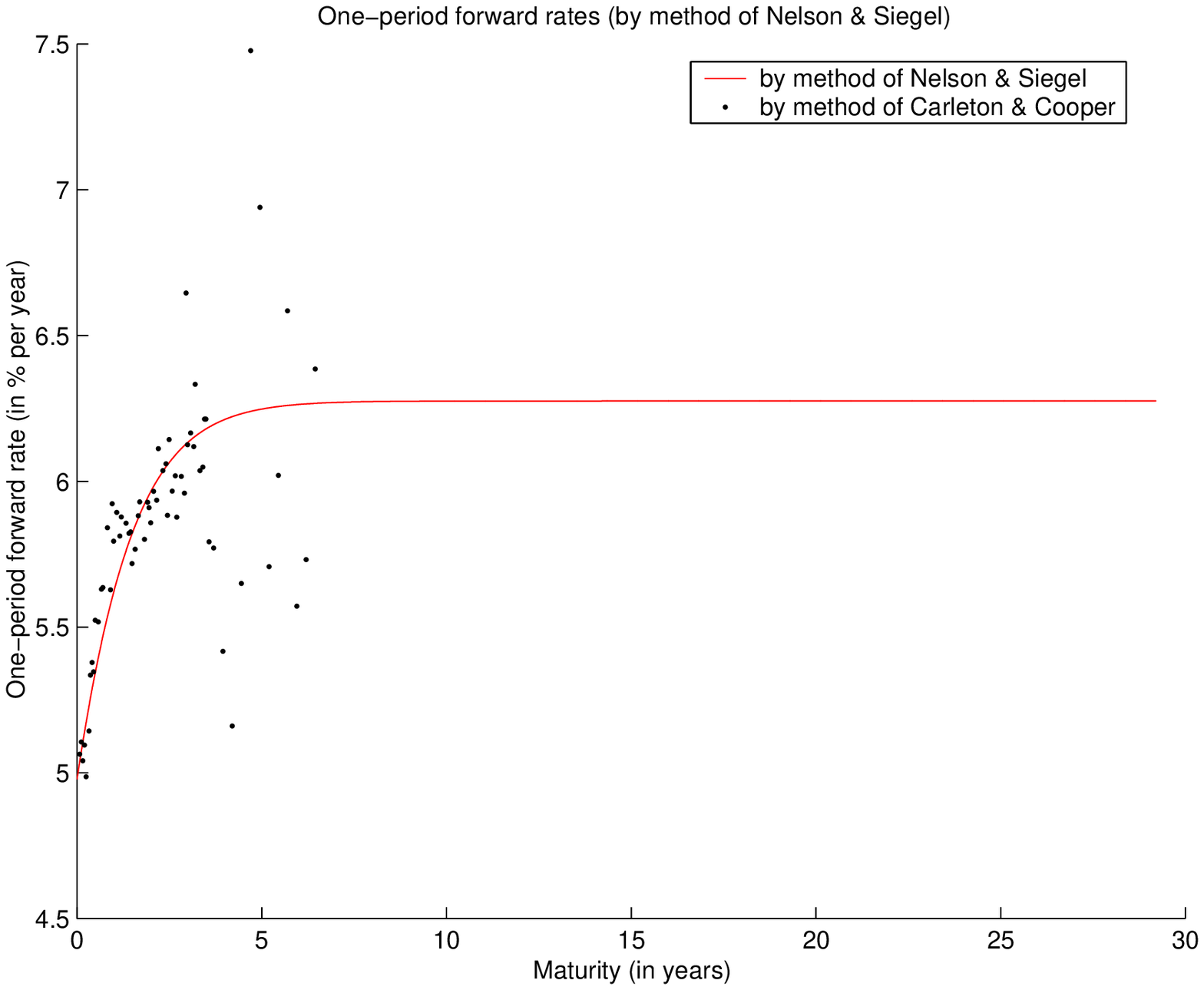} 
\end{center} 
\caption{\label{fig:Nelson}} 
\end{figure}

The graphs in figure \ref{fig:Nelson} are produced following the 
approach of Nelson and Siegel. The main features of the estimates 
based on parsimonious models are readily seen. Although for the 
discount function the differences with other approaches are pretty 
mild, the constraints imposed on the functional form of the 
instantaneous forward rate function get rid of the oscillating 
behavior peculiar to spline based techniques. 
 
We conclude this section by showing a table where the Root Mean 
Square Error (RMSE) 
\footnote{{\it i.e.} the square root of the arithmetic mean of 
squared residuals, see equation \eqref{eq:minimizzazione}} 
obtained in our numerical experiments is reported. This allows us 
to compare the quality of the fit among all employed  methods (for various 
choices of the parameters, such as knot points or degree). 
 
\begin{table} 
\begin{center} 
\begin{tabular}{| l | l | c | c |} 
        \hline 
        Method    & Knot points           & spline degree & RMSE  \\ \hline 
        \hline \\ 
        Carleton  &                       &               & .0807 \\ \hline 
        \hline \\ 
        McCulloch & {\it ala} McCulloch   & 2             & .2428 \\ \hline 
        McCulloch & {\it ala} McCulloch   & 3             & .2441 \\ \hline 
        McCulloch & {\it ala} McCulloch   & 4             & .1896 \\ \hline 
        \hline \\ 
        McCulloch & 0, 1, 3, 5, 7, 11, 30 & 2             & .2493 \\ \hline 
        McCulloch & 0, 1, 3, 5, 7, 11, 30 & 3             & .2358 \\ \hline 
        McCulloch & 0, 1, 3, 5, 7, 11, 30 & 4             & .2389 \\ \hline 
        McCulloch & 0, 1, 3, 5, 7, 11, 30 & 5             & .1846 \\ \hline 
        \hline \\ 
        Coleman   & {\it ala} McCulloch   & 0             & .3003 \\ \hline 
        Coleman   & {\it ala} McCulloch   & 1             & .2350 \\ \hline 
        Coleman   & {\it ala} McCulloch   & 2             & .2328 \\ \hline 
        Coleman   & {\it ala} McCulloch   & 3             & .1926 \\ \hline 
        \hline \\ 
        Coleman   & 0, 1, 3, 5, 7, 11, 30 & 0             & .4560 \\ \hline 
        Coleman   & 0, 1, 3, 5, 7, 11, 30 & 1             & .2242 \\ \hline 
        Coleman   & 0, 1, 3, 5, 7, 11, 30 & 2             & .2405 \\ \hline 
        Coleman   & 0, 1, 3, 5, 7, 11, 30 & 3             & .2282 \\ \hline 
        \hline \\ 
        Nelson    &                       &               & .5102 \\ \hline 
\end{tabular} 
\end{center} 
\caption{Fit quality} 
\end{table} 
 
As expected the quality of the fit is higher for spline based
techniques and lower for parsimonious models.  Moreover it is worth
noting that it is not true that increasing the order of the spline or
the number of knots yields a better fit.

\section{Conclusions \label{sect:conclusioni}}
\noindent We have presented a critical review of the main methods for
the estimation of the term structure of interest rates in fixed income
markets.

It is apparent that most of the activity in this field lacks of a
solid mathematical basis and there is not a single technique which is
definitely {\it better} than others.

A number of elements contribute to form this scenario:
\begin{trivlist}
    \item [Bad practice:]  there is a general trend of re-cycling
existing techniques without any critical evaluation of their real
effectiveness.
    \item [Low-quality data:]  errors can be present in the input data
(wrong prices).  A large error is immediately detected whereas smaller
errors may induce to take into account ``fake'' arbitrage
opportunities.
    \item [Differences among the markets:]  fixed income markets are
not homogeneous.  The number of data available for a cross-section
analysis of the U.S.  Treasury securities is very different, for
instance, from the number of outstanding Italian government bonds.
The features of a market should be considered very carefully before
applying any technique.
\end{trivlist}

We expect to extend our work in two directions:
\begin{trivlist}
    \item The development of algorithms to find the best position of
the knots for spline-based interpolation methods.  We consider genetic
algorithms very promising for this kind of problem.
    \item A mathematical analysis of equilibrium term structure
models.  This is pretty important since there is already evidence of
relevant differences between the predictions of existing models and
the experimental data (an example is the volatility term structure).
\end{trivlist}

The importance of ``intuition'' in financial markets analysis cannot
be underestimated.  However a more precise distinction among
hypotheses, mathematical implications of these hypotheses and
empirical observations is definitely required in this field.
 
\section{Appendix} 
 
This appendix is entirely devoted to the proof of Theorem 
\ref{thm:fattori di sconto} that we write again here for our 
convenience. 
 
\begin{theorem}
    Let $\T\subset\B$ be a complete coupon term structure and let 
$T(\T)=(t_1,\ldots,t_N)$ for some $N\in\N$.  If $\Cardinality(\T) 
\geq N$ and if the following condition: 
\begin{enumerate} 
    \ipotesi{$(NA_1)$} 
if $(q_1,\ldots,q_B)\in\R^B$ are such that $\sum_{j=1}^B q_j 
\boldsymbol{\varphi}(c_j,m_j)=0$, then $\sum_{j=1}^B q_j p_j=0$; 
\end{enumerate} 
is satisfied, then there exist $d_1,\ldots,d_N\in\R$ such that for 
every $(c,m,p)\in\T$ 
\begin{equation}\label{eq:PREZZO} 
    p=\sum_{i=1}^N d_i \varphi_i(c,m). 
\end{equation} 
If, furthermore, conditions: 
\begin{enumerate} 
    \ipotesi{$(NA_2)$}
if $(q_1,\ldots,q_B)\in\R^B$ is such that
\begin{equation*}
    \sum_{j=1}^B q_j \varphi_i(c_j,m_j) =
    \begin{cases}
        f_{\bar{\imath}} &, \quad\text{if}\; i =\bar{\imath} \\
        0                &, \quad\text{otherwise}
    \end{cases},
\end{equation*} 
for some $f_{\bar{\imath}} >0$, then $0 <\sum_{j=1}^B q_j p_j
<f_{\bar{\imath}}$;
    \ipotesi{$(NA_3)$} 
if $(q_1,\ldots,q_B)\in\R^B\backslash\{0\}$ are such that 
$\sum_{j=1}^B q_j p_j=0$, then there exists 
$\bar{\imath}\in\{1,\ldots,N\}$ such that $\sum_{i=1}^{\bar{\imath}} 
\sum_{j=1}^B q_j\varphi_i(c_j,m_j)\geq 0$; 
\end{enumerate} 
are fulfilled as well, then $1>d_1>d_2>\ldots>d_N>0$. 
\end{theorem} 
 
\begin{proof} 
    Let $(c_1,m_1,p_1),\ldots,(c_N,m_N,p_N)\in\T$ be such that 
$m_j=t_j$ for every $j\in\{1,\ldots,N\}$. The existence of $N$ 
bonds like that in the set $\T$ is ensured by the hypotheses that 
$\Cardinality(\T) \geq N$ and that $\T$ forms a complete coupon 
term structure. 
 
Let $\Phi=((\Phi_{ij}))$ be an $N\times N$ matrix such that 
\begin{equation}\label{eq:matrice dei cash-flows} 
    \Phi_{ij} :=\varphi_i(c_j,m_j) 
\end{equation} 
for all $i, j\in\{1,\ldots,N\}$. 
 
Finally let $(c,m,p)$ be an arbitrary bond of $\T$. 
 
Since $\Phi$ is an upper triangular matrix with no vanishing 
element on its principal diagonal, it is non-singular. 
 
Now let $q_1,\ldots,q_N\in\R$ be such that 
\begin{equation*} 
    q_j =\sum_{i=1}^N (\Phi^{-1})_{ji} \varphi_i(c,m) \qquad (j=1,\ldots,N) 
\end{equation*} 
Then 
\begin{equation*} 
    \sum_{j=1}^N q_j \boldsymbol{\varphi}(c_j,m_j) 
=\boldsymbol{\varphi}(c,m). 
\end{equation*} 
By condition \ref{cond:arbitraggio 1}, this implies equation
\eqref{eq:PREZZO} with
\begin{equation*} 
    d_i :=\sum_{j=1}^N p_j (\Phi^{-1})_{ji} \qquad (i=1,\ldots,N). 
\end{equation*} 
In order to conclude the proof of the first part of the theorem, 
we have to show that the $d_i$'s are independent of the choice of 
$(c_1,m_1,p_1),\ldots,(c_N,m_N,p_N)$ such that $m_j=t_j$ for every 
$j$ if several such choices are possible. To do this, let 
$\bar{\jmath}\in\{1,\ldots,N\}$ and let 
$(\tilde{c}_{\bar{\jmath}}, \tilde{m}_{\bar{\jmath}}, 
\tilde{p}_{\bar{\jmath}}) \in\T\backslash\{(c_{\bar{\jmath}}, 
m_{\bar{\jmath}}, p_{\bar{\jmath}})\}$ be such that 
$\tilde{m}_{\bar{\jmath}} =t_{\bar{\jmath}}$. 
 
Let $\tilde{\Phi}=((\tilde{\Phi}_{ij}))$ be an $N\times N$ matrix 
such that 
\begin{equation}\label{eq:nuova matrice dei cash-flows} 
    \tilde{\Phi}_{ij} := 
    \begin{cases} 
        \varphi_i(c_j,m_j) &, \quad\text{if}\; j \neq\bar{\jmath} 
\\ 
        \varphi_i(\tilde{c}_{\bar{\jmath}}, 
\tilde{m}_{\bar{\jmath}})  &, \quad\text{if}\; j =\bar{\jmath} 
    \end{cases} 
\end{equation} 
for all $i, j\in\{1,\ldots,N\}$, and let 
\begin{equation*} 
    \tilde{d}_i :=\sum_{\stackrel{j=1}{j \neq\bar{\jmath}}}^N p_j 
(\tilde{\Phi}^{-1})_{ji} +\tilde{p}_{\bar{\jmath}} 
(\tilde{\Phi}^{-1})_{\bar{\jmath} i} \qquad (i=1,\ldots,N). 
\end{equation*} 
We want to show that $\tilde{d}_i =d_i$ for every 
$i\in\{1,\ldots,N\}$. Since both $\bar{\jmath}$ and 
$(\tilde{c}_{\bar{\jmath}}, \tilde{m}_{\bar{\jmath}}, 
\tilde{p}_{\bar{\jmath}})$ has been arbitrarily chosen, this will 
conclude the proof of the first part of the theorem. 
 
Observe that $\tilde{\Phi}_{\cdot j} =\Phi_{\cdot j}$ for all $j 
\neq\bar{\jmath}$ and that by equations \eqref{eq:PREZZO}, 
\eqref{eq:matrice dei cash-flows} and \eqref{eq:nuova matrice dei 
cash-flows} 
\begin{equation*} 
    p_j =\sum_{i=1}^N d_i \Phi_{ij} \qquad (j=1,\ldots,N) 
\end{equation*} 
and 
\begin{equation*} 
    \tilde{p}_{\bar{\jmath}} =\sum_{i=1}^N d_i \tilde{\Phi}_{i 
\bar{\jmath}}. 
\end{equation*} 
Then for every $i\in\{1,\ldots,N\}$ 
\begin{equation*} 
    \begin{align*} 
        \tilde{d}_i 
&=\sum_{\stackrel{j=1}{j \neq\bar{\jmath}}}^N 
\left[(\tilde{\Phi}^{-1})_{ji} \sum_{h=1}^N d_h \Phi_{hj}\right] + 
(\tilde{\Phi}^{-1})_{\bar{\jmath} i} \left(\sum_{h=1}^N d_h 
\tilde{\Phi}_{h \bar{\jmath}}\right) \\ 
        &=\sum_{h=1}^N d_h \left[\sum_{j=1}^N 
\tilde{\Phi}_{hj}(\tilde{\Phi}^{-1})_{ji}\right] =d_i. 
    \end{align*} 
\end{equation*} 
 
In order to prove the second part of the theorem, let
$(c_1,m_1,p_1),\ldots,(c_N,m_N,p_N)\in\T$ be such that $m_j=t_j$ for
every $j\in\{1,\ldots,N\}$ and let $(q_1,\ldots,q_N) \in\R^N$ be a
portfolio of these bonds such that
\begin{equation*} 
    \sum_{j=1}^N q_j \varphi_i(c_j,m_j) = 
    \begin{cases} 
        f_{\bar{\imath}}   &, \quad\text{if}\; i=\bar{\imath} \\ 
        f_{\bar{\imath}+1} &, \quad\text{if}\; i=\bar{\imath}+1 \\ 
        0                  &, \quad\text{otherwise} 
    \end{cases}, 
\end{equation*} 
for some $\bar{\imath}\in\{1,\ldots,N\}$ and some $f_{\bar{\imath}},
f_{\bar{\imath}+1} \in\R$.  By \eqref{eq:PREZZO}, the price of this
portfolio is
\begin{equation*} 
    \sum_{j=1}^N q_j p_j =d_{\bar{\imath}} f_{\bar{\imath}}
+d_{\bar{\imath} +1} f_{\bar{\imath} +1}.
\end{equation*}
If $f_{\bar{\imath}} >0$ and $f_{\bar{\imath} +1} =0$, then by
condition \ref{cond:arbitraggio 2}, $0 <d_{\bar{\imath}}
f_{\bar{\imath}} <f_{\bar{\imath}}$, which implies that $0
<d_{\bar{\imath}} <1$ and, since $\bar{\imath}$ has been chosen
arbitrarily, that
\begin{equation*} 
    0<d_i<1 \qquad (i=1,\ldots,N).
\end{equation*}

Now suppose, by contradiction, that $d_{\bar{\imath}}\leq
d_{\bar{\imath}+1}$ and let $f_{\bar{\imath}}, f_{\bar{\imath}+1}$ be
such that
\begin{equation*} 
    d_{\bar{\imath}} f_{\bar{\imath}} +d_{\bar{\imath}+1}
f_{\bar{\imath}+1} =\sum_{j=1}^N q_j p_j =0.
\end{equation*}
Since $d_{\bar{\imath}}, d_{\bar{\imath}+1}
>0$, we can assume, without loss of generality, that $f_{\bar{\imath}}
\geq 0$ and that $f_{\bar{\imath}} +f_{\bar{\imath}+1} \geq 0$.
However condition \ref{cond:arbitraggio 3} implies that either
$f_{\bar{\imath}} <0$ or $f_{\bar{\imath}} +f_{\bar{\imath}+1} <0$.
This is the wanted contradiction.
\end{proof}

 
\newcommand{\noopsort}[1]{}
\providecommand{\bysame}{\leavevmode\hbox to3em{\hrulefill}\thinspace}

\end{document}